\documentclass[conference]{IEEEtran}
\usepackage{cite}
\usepackage{bm}
\usepackage{subfig}
\usepackage{caption,xfrac,amssymb}
\usepackage[font=footnotesize]{caption}
\usepackage[]{graphicx}
\usepackage{siunitx}
\usepackage{balance}
\ifCLASSINFOpdf
\else
\fi
\usepackage{url}

\usepackage[]{fancyhdr} %
\newcommand{\changefont}{\fontsize{7}{7}\selectfont}

\DeclareFontFamily{U}{matha}{\hyphenchar\font45}
\DeclareFontShape{U}{matha}{m}{n}{
      <5> <6> <7> <8> <9> <10> gen * matha
      <10.95> matha10 <12> <14.4> <17.28> <20.74> <24.88> matha12
      }{}
\DeclareSymbolFont{matha}{U}{matha}{m}{n}

\DeclareMathSymbol{\Lt}{3}{matha}{"CE}
\DeclareMathSymbol{\Gt}{3}{matha}{"CF}

\newcommand \bV{\mathbf{V}}
\newcommand \bY{\mathbf{Y}}

\newcommand \bB{\mathbf{B}}
\newcommand \bG{\mathbf{G}}
\newcommand \bI{\mathbf{I}}
\newcommand \bi{\mathbf{i}}
\newcommand \bF{\mathbf{F}}
\newcommand \bL{\mathbf{L}}
\newcommand \bR{\mathbf{R}}
\newcommand \bP{\mathbf{P}}
\newcommand \bQ{\mathbf{Q}}
\newcommand \bS{\mathbf{S}}

\newcommand \bx{\mathbf{x}}
\newcommand \be{\mathbf{e}}
\newcommand \by{\mathbf{y}}
\newcommand \bv{\mathbf{v}}

\newcommand \bE{\mathbf{E}}
 
\newcommand \bof{\mathbf{f}}
\newcommand \bg{\mathbf{g}}

\newcommand \diag{\mathrm{diag}}

\newcommand \bchi{\boldsymbol{\chi}}

\newcommand \mra{\mathrm{a}}
\newcommand \mre{\mathrm{e}}
\newcommand \mrs{\mathrm{s}}
\newcommand \mrss{\mathrm{ss}}
\newcommand \bdelta{\boldsymbol{\delta}}
\newcommand \bomega{\boldsymbol{\omega}}
\newcommand \bone{\mathbf{1}}
\newcommand \bzero{\mathbf{0}}
\newcommand \bGamma{\boldsymbol{\Gamma}}

\newcommand \mcG{\mathcal{G}}
\newcommand \mcN{\mathcal{N}}
\newcommand \mcI{\mathcal{I}}
\newcommand \mcE{\mathcal{E}}

\usepackage[thinc]{esdiff}
\usepackage{amsmath,bbm,dsfont}
\usepackage{color}
\usepackage{booktabs}
\usepackage{threeparttable, ltablex}

\usepackage{cite}
\IEEEoverridecommandlockouts

\usepackage{authblk}

\allowdisplaybreaks
\begin{document}

%
\title{Integrated System Models for Networks with Generators \& Inverters\thanks{Funding provided by the U.S. Department of Energy (DOE) Office of Energy Efficiency and Renewable Energy under Solar Energy Technologies Office (SETO) through the award numbers EE0009025 and 38637 (UNIFI consortium), respectively; from the National Science Foundation grant 1444745; and University of Minnesota's MnDRIVE program is also gratefully acknowledged. The views expressed herein do not necessarily represent the views of the U.S. Department of Energy or the United States Government. \\$^\ddagger$D. Venkatramanan and M. K. Singh contributed equally to the paper.}}



\author[$\dagger \ddagger$]{D. Venkatramanan}
	\author[$\dagger \ddagger$]{Manish K. Singh}
	\author[$\star$]{Olaoluwapo Ajala}
	\author[$\star$]{Alejandro Dom\'inguez-Garc\'ia}
	\author[$\dagger$]{Sairaj Dhople}
	\affil[$\dagger$]{\small Department of Electrical and Computer Engineering, University of Minnesota, Minneapolis, MN, USA}
	\affil[$\star$]{\small Department of Electrical and Computer Engineering, University of Illinois, Urbana, IL, USA}
\affil[ ]{E-mails: dvenkat@umn.edu, mksingh@umn.edu, ooajala2@illinois.edu, aledan@illinois.edu, sdhople@umn.edu}


%





\maketitle
\thispagestyle{fancy}
\pagestyle{fancy}

\begin{abstract}
Synchronous generators and inverter-based resources are complex systems with dynamics that cut across multiple intertwined physical domains and control loops. Modeling individual generators and inverters is, in itself, a very involved activity and has attracted dedicated attention from power engineers and control theorists over the years. Control and stability challenges associated with increasing penetration of grid-following inverters have generated tremendous interest in grid-forming inverter technology. The envisioned coexistence of inverter technologies alongside rotating machines call for modeling frameworks that can accurately describe networked dynamics of interconnected generators and inverters across timescales. We put forth a comprehensive integrated system model for such a setting by: i)~adopting a combination of circuit- and system-theoretic constructs, ii)~unifying representations of three-phase signals across reference-frame transformations and phasor types, and iii)~leveraging domain-level knowledge, engineering insights, and reasonable approximations. A running theme through our effort is to offer a clear distinction between physics-based models and the task of modeling. Among several insights spanning the spectrum from analytical to practical, we highlight how differential-algebraic-equation models and algebraic power-flow phasor models fall out of the detailed originating electromagnetic transient models. 
\end{abstract}


\IEEEpeerreviewmaketitle

\section{Introduction}
\begin{figure*}
			\centering
			{\includegraphics[width = 1\linewidth]{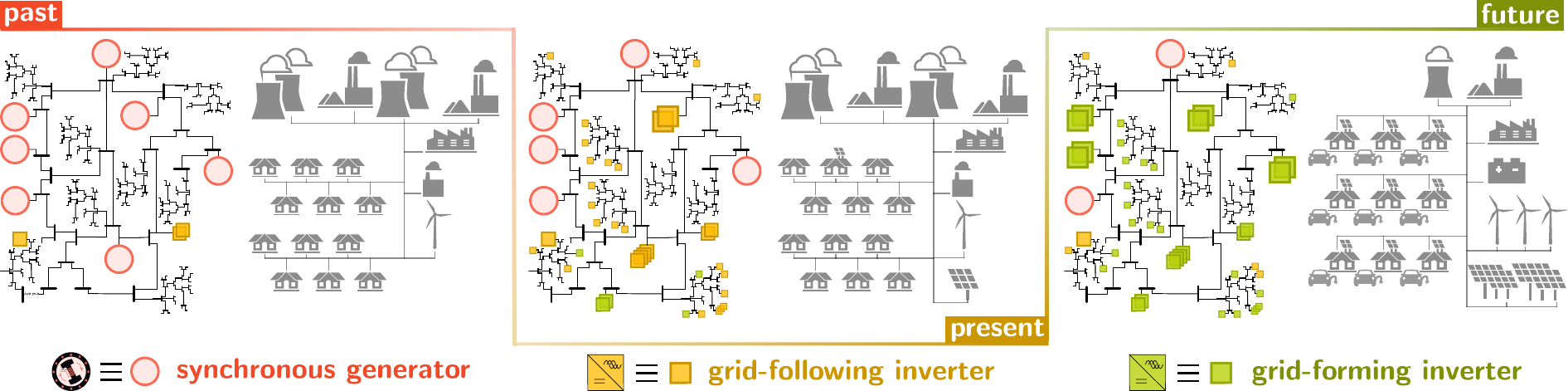}} 
	        \caption{{Schematic illustration of the transformational change taking place in the generation mix of the electric power grid.}}
	\label{Fig:ppp}
\end{figure*}

This paper establishes an integrated system model for electrical networks composed of: i)~three-phase synchronous generators (SGs); ii)~grid following (GFL) inverter-based resources (IBRs) with synchronous reference-frame phase-locked loops (PLLs); and iii)~droop, virtual synchronous machine (VSM), and virtual oscillator controlled (VOC) grid-forming (GFM) IBRs. A unified representation for interconnections of these heterogeneous resources is achieved by leveraging a circuit-theoretic exposition. In particular, we put forth dependent voltage-source models for generators and inverters with network-facing electrical quantities (voltage magnitudes, phase angles, frequencies) assuming a consistent and continuous interpretation across resources and timescales. These quantities---critical as they may be to grid operations, analysis, and control---have been shrouded behind a thin veil of ambiguity when presented collectively in previous works. The modeling formulation we offer dispels  ambiguities that are particularly glaring when resources of different types are all modeled on the same page. Further, we project that the unified perspective we offer will fuel the ongoing development of control schemes, optimization algorithms, stability-assessment methods, and simulation tools for grids with a heterogeneous mix of generators and inverters~\cite{Denis-2018,Markovic21SSstability,SKL19GM,Sanjana21GM,Tayyebi20FreqStability,Rodrigo21Magazine,Duncan-2021}. There is a broad-based recognition that IBRs of different types (GFM and GFL) will co-exist alongside generators for many years to come~\cite{Taylor-2016-Fuel,Milano18foundation}; {Fig.~\ref{Fig:ppp} illustrates the transformational change taking place in the generation mix}. Modeling and control efforts are underway to facilitate different stages of the ongoing compositional shift in the power industry~\cite{TG22Mapping}. The setting we examine is motivated by and structured to translate benefits to these endeavors. 
Contemporary research in inverter-dominant grids oftentimes blurs the distinction between ground-truth models, representation of ground-truth models in various formats, and modeling approximations. This is attributable to several factors that we overview below:
\begin{itemize}
\renewcommand{\labelitemi}{\tiny$\blacksquare$}
    \item Power-electronics engineers who have been trained in and used to model and control single-inverter units have recently been facing the daunting challenge of translating their models and designs to networks of large collections of inverters~\cite{SCHIFFER16survey}.
    \item Power-systems engineers have had to acknowledge inverter dynamics in several tasks pertinent to grid modeling, simulation, and operations. Dynamics of IBRs are dominated by those of the companion controllers; with myriad control schemes all fighting for attention across industry and academia.
    \item Simulation tools for the integrated modeling of generators and inverters are in a state of infancy; there is limited attention devoted to appreciating the difference between how controllers are realized and how they are represented in models.
    \item Pertinent phenomena and dynamics are to be modeled over wide timescales (real-time through to sinusoidal steady state), and multiple physical domains underline the properties of the resources (electrical and mechanical, and coupled electromagnetic and electromechanical).
\end{itemize}

Our effort addresses the challenges listed above from several angles. The models are stitched together in a manner that clearly highlights how electromechanical physics dominates the dynamics of synchronous generators, while controls dominate the dynamics of inverters. The language of circuits provides a unifying thread to tie together electrical signals across the resources and across timescales. We make a concerted effort to clarify what facets of the presented resource and network dynamics originate from physics and can be justified based on first principles, and what facets are to be appreciated as modeling effort(s). The paper also provides significant pedagogical value; since, in the process of setting up the models, we provide a comprehensive discussion on reference-frame transformations, phasors (of different types), and representations of electrical signals and notions of power in different reference frames. 
A summary of contributions is provided below: 
\begin{itemize}
\renewcommand{\labelitemi}{\tiny$\blacksquare$}
    \item Integrated system models presented in precise circuit-theoretic terms with voltage-behind-$RL$ (under dynamic conditions) and voltage-behind-reactance (under steady state) models for generators and inverters.  
    \item Cohesive presentation of models in $\mathrm{abc}$, direct-quadrature, and sinusoidal steady-state phasor regimes. Translations across regimes accomplished without loss of clarity or obfuscation of involved signals. 
    \item Catalog of full suite of models while narrating the origin stories for approximate models (power flow, differential algebraic equation (DAE)) that are commonly referenced in the literature. 
    \item Unified presentation of models for generators and inverters in a manner that, in the electrical domain, retains the same meaning for angles, frequencies, terminal voltages. 
    \item Definitive guide to representation of three-phase signals in different reference frames with the bare minimum assumptions on sinusoidal form. Clear connections across reference frames established.
\end{itemize}
{Via the contributions summarized above, we dwell on the following intriguing questions that have not received rigorous attention in the literature, in spite of being frequently echoed in conversations on topics pertinent to modeling, analysis, and control of networked generator and inverter resources: 
\begin{itemize}
    \item[$\mathcal{Q}1$)] Are GFL IBRs current sources and GFM IBRs voltage sources? What does that mean in precise circuit- and system-theoretic contexts?
    \item[$\mathcal{Q}2$)] What is the precise relationship between true-to-form dynamic models in stationary $\mathrm{abc}$ frame, magnitude and phase of complex-valued local direct-quadrature ($\mathrm{dq}$), global direct-quadrature $\mathrm{DQ}$, and sinusoidal steady-state phasors?
    \item[$\mathcal{Q}3$)] Can we rigorously obtain (or justify) the steady-state models for SGs, GFL IBRs, and GFM IBRs for power-flow formulations?
    \item[$\mathcal{Q}4$)] How does the difference between electrical-radian synchronous frequency and steady-state frequencies that networks actually operate at manifest in models across timescales? 
    \item[$\mathcal{Q}5$)] What precise simplifications of original true-to-form dynamic models for resources and networks yield differential algebraic equation (DAE) models and steady-state power-flow models?
    \item[$\mathcal{Q}6$)] What is the difference between modeling (to facilitate simulation with computing ease) and models (grounded in first-principles electromagnetic and electromechanical theory) vis-\`a-vis synchronous generators, GFL IBRs, and GFM IBRs? 
\end{itemize}
}
In pointing out the scope of the paper, what it covers, and what its contributions are, it is equally important to highlight what facets are out of scope. Particularly, the paper does not foray into the following aspects:
\begin{itemize}
\renewcommand{\labelitemi}{\tiny$\blacksquare$}
    \item Simulation methods and tools (commercial off-the-shelf, open-source, and others) and their suitability for the different types of models. 
    \item Approaches (analytical or computational) to contend with model complexity (e.g., model-reduction methods). 
    \item Secondary and tertiary control strategies to restore frequency to synchronous and optimally operate networks based on power quality, cost, and fuel considerations. 
\end{itemize}

We review closely related previous efforts that have attempted to provide unified and/or circuit-based models of individual resources and/or integrated systems and place the contributions of our work in context. There have been attempts to: unify models for different types of GFMs~\cite{Ajala22HiCSS}, examine GFL models with a circuit-theoretic lens~\cite{Minghui2021hicss,DR1}, and reveal the relationship between GFM control types~\cite{AreSuul14Equivalent,Sinha-2017}. These efforts, however, do not present models for all three types of resources we consider in this work (SGs, GFM IBRs, GFL IBRs) on a common platform. Assumptions and modeling approximations made for one type of resource (or a system composed of dominantly one type of resource) do not hold in general across the board in heterogeneous settings. This can render models derived in isolation to be inaccurate, particularly with regard to capturing the dynamical behavior of networked resources in complex systems. Other attempts in~\cite{Green-2021a, Green-2022a} employ an impedance-based approach using transfer functions for modeling resources in a composite-network setting and examining negative resource interactions and stability at large. However, a systematic effort to place all three types of resources on the same canvas with precise elucidation of full-order dynamic models, accounting for all device- and network-facing parameters and signals has been lacking in all these previous efforts. With regard to reviewing related literature that is aligned from a system-theoretic standpoint, we point to the closely related effort~\cite{SCHIFFER16survey}. In reference~\cite{SCHIFFER16survey}, prevalent reference-frames are reviewed and two broad model classes are elaborated upon: the first models network dynamics and generic inverter dynamics invoking $\mathrm{abc}$ reference frame for inverters; the second ignores network and GFL dynamics and uses direct-quadrature representation to model generic GFM dynamics. While the previous work focuses on IBRs, we include synchronous generator models alongside IBRs, all presented in a unified circuit-theoretic interpretation. Moreover, we provide detailed network and resource dynamics in both $\mathrm{abc}$ and direct-quadrature representations which culminate to a steady-state integrated system model with phasor representations. The exposition in~\cite{SCHIFFER16survey}, at an early stage, assumes three-phase sinusoidal signals with identical amplitudes and $120^\circ$ phase-shift. However, in our work it is shown that elaborate dynamic models of varying complexities can indeed be developed with minimal assumption on the grid signals. We assume the three-phase signals to be sinusoidal with common amplitude and $120^\circ$ phase-shift only while considering steady-state operation. In summary, compared to~\cite{SCHIFFER16survey}, this work includes greater detail on resource models, includes synchronous generators, and provides a full suite of models in prevalent reference frame representations at different time scales. 

The remainder of this paper is organized as follows. Section~\ref{sec:prelim} overviews mathematical preliminaries and associated fundamentals related to signal representation across reference frames. An overview of the integrated system modeling formulations is provided in Section~\ref{sec:GenModels}. Network-level models in $\mathrm{abc}$ reference frame, the global $\mathrm{DQ}$ reference frame, and sinusoidal steady state are discussed in Section~\ref{sec:Network}. In Section~\ref{sec:Resource}, we discuss the dynamic models for the generators and IBRs; we also outline abstractions that are relevant in steady state. Inferences and offshoots that follow from the integrated system models are discussed in Section~\ref{sec:ICO}. Concluding remarks are presented in Section~\ref{sec:conclude}.

\section{Preliminaries} \label{sec:prelim}
In this section, we establish notation; discuss the different reference frames and phasor types for signal representation; and finally, comment on notions of active, reactive, and complex-apparent power in the different reference frames. 

\subsection{Notation} 
The matrix transpose will be denoted by $(\cdot)^\top$, complex conjugate by $(\cdot)^*$, real and imaginary parts of a complex number by $\Re(\cdot)$ and $\Im(\cdot)$, respectively, magnitude of a complex scalar by $|\cdot|$, and $\jmath = \sqrt{-1}$. Operator $\circ$ denotes entry-wise product for vectors. Boldface symbols denote matrices and vectors. A diagonal matrix formed with entries of a vector $\bx$ is denoted by $\diag(\bx)$. For vector $\bx$ with $n$-th entry $x_n$, $\sin(\bx)$ and $\cos(\bx)$ are vectors with $n$-th entries $\sin(x_n)$ and $\cos(x_n)$, respectively. For scalar $\theta$, boldface $\boldsymbol{\sin}()$ and $\boldsymbol{\cos}()$ are defined as
\begin{align*}
    \boldsymbol{\sin}(\theta) &= \begin{bmatrix}\sin \theta&
    \sin\big(\theta-\frac{2\pi}{3}\big)&
    \sin\big(\theta+\frac{2\pi}{3}\big)\\
    \end{bmatrix}^\top,\\
    \boldsymbol{\cos}(\theta) &= \begin{bmatrix}\cos \theta&
    \cos\big(\theta-\frac{2\pi}{3}\big)&
    \cos\big(\theta+\frac{2\pi}{3}\big)\\
    \end{bmatrix}^\top.
\end{align*}
The space of $M\times N$ real-valued matrices is denoted by $\mathds{R}^{M\times N}$. With $\bzero$ and $\bone$, we denote column vectors (of appropriate length) with all entries equal to $0$ and $1$, respectively. The $N\times N$ identity matrix is denoted by $\mathbbm{I}_N$. We will also utilize the following matrix to aid exposition: 
     $\boldsymbol{\Upsilon} = \frac{2}{3}\begin{bmatrix} 0 & -1\\ 1 & 0\end{bmatrix}$.

\subsection{Reference Frames \& Phasor Types}
\subsubsection{Three-phase $\mathrm{abc}$ Signals and Space Phasors}
We begin our exposition with three-phase time-domain signals collected in $\bx = [x_\mathrm{a}, x_\mathrm{b}, x_\mathrm{c}]^\top$. We focus our analysis on systems where all signals (imposed and realized) sum to zero, i.e., $$x_\mathrm{a} + x_\mathrm{b}+ x_\mathrm{c}=0.$$ These are called \textit{balanced signals}. We do not ascribe the signals any particular form (e.g., sinusoidal) in this representation. Notice that the zero-sum 3-phase signals can be expressed using two independent variables. One such prevalent representation is termed as the \textit{space phasor}~\cite{Iravani_Book10} (also known as \textit{space vector} in power-electronics and electric-drives parlance~\cite{Lipo2003pulse}), which is defined as a complex-valued scalar: 
\begin{equation}\label{eq:SPdef}
    \overrightarrow{X} = [1, \mathrm{e}^{\jmath \frac{2\pi}{3}}, \mathrm{e}^{\jmath \frac{4\pi}{3}}] \bx.
\end{equation} 
Decomposing $\overrightarrow{X} = \Re(\overrightarrow{X}) + \jmath \Im(\overrightarrow{X})$, we can write:
\begin{align}\label{eq:phasor-ABC}
    \begin{split}
        \begin{bmatrix} \Re(\overrightarrow{X}) \\ \Im(\overrightarrow{X}) \end{bmatrix} = \begin{bmatrix}1 & -\frac{1}{2} & -\frac{1}{2}\\ 0 & \frac{\sqrt{3}}{2} &-\frac{\sqrt{3}}{2} \end{bmatrix} \begin{bmatrix}x_\mathrm{a} \\ x_\mathrm{b} \\ x_\mathrm{c}\end{bmatrix}=\bGamma\begin{bmatrix}x_\mathrm{a} \\ x_\mathrm{b} \\ x_\mathrm{c}\end{bmatrix}.
    \end{split}
\end{align}
In the other direction, we can show that the $\mathrm{abc}$ signals, $x_\mathrm{a}, x_\mathrm{b}, x_\mathrm{c}$, are given by
\begin{align}\label{eq:ABC-phasor}
    \begin{split}
        \begin{bmatrix}x_\mathrm{a} \\ x_\mathrm{b} \\ x_\mathrm{c}\end{bmatrix} = \frac{2}{3}\bGamma^\top \begin{bmatrix} \Re(\overrightarrow{X}) \\ \Im(\overrightarrow{X}) \end{bmatrix}.
    \end{split}
\end{align}


\subsubsection{Direct-quadrature Reference Frames}
Recognizing that power-system quantities are typically close to sinusoidal signals, it is often convenient to project the space phasors on to rotating reference frames (with direct and quadrature components) for ease of analysis. Such representations yield the benefits of having:~i)~a decoupled set of dynamical model equations for three-phase systems (i.e., with no mutual coupling terms involved), ii)~time-invariant system parameters (e.g., in synchronous machines models), and iii)~dc signals in steady state (that allows use of traditional PI controllers in the control layer to achieve zero steady-state errors). One option is to select the reference frames for all signals (voltages and currents across the network) to be at a common constant rotation frequency, typically, the \emph{electrical radian synchronous frequency} $\omega_\mathrm{s} = 2 \pi f_\mathrm{s}$ (where $f_\mathrm{s} = 50$ or $60$~$\mathrm{Hz}$). We denote such a projection by complex-valued signal in polar form, $X$: 
\begin{equation}\label{eq:DQintro}
    X = |X|\mathrm{e}^{\jmath\rho},
\end{equation}
and point out the following relationship to the originating space phasor: 
\begin{align} \label{eq:DQdef}
\begin{split}
    \overrightarrow{X} &= X \mathrm{e}^{\jmath \omega_\mathrm{s}t} = |X|  \mathrm{e}^{\jmath (\omega_\mathrm{s} t + \rho)}.
    \end{split}
\end{align}

\begin{figure}
			\centering
			{\includegraphics[width = 1\linewidth]{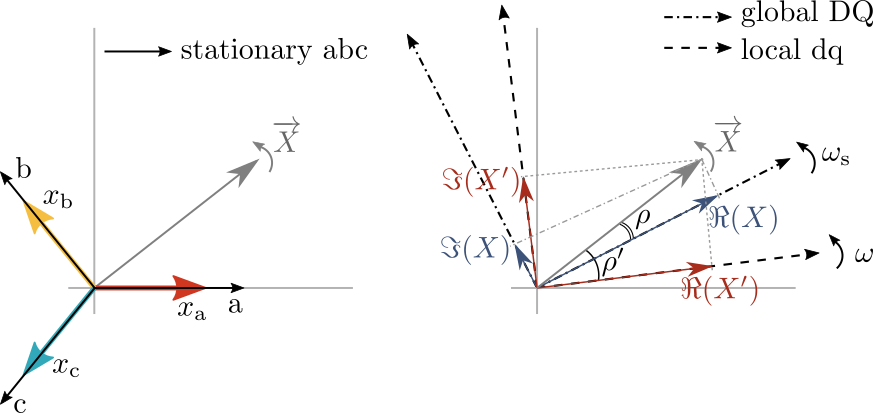}} 
	        \caption{(a) Space phasor shown in (stationary) $\mathrm{abc}$ reference frame,and (b) projection of the space phasor on (rotating) global $\mathrm{DQ}$ and local $\mathrm{dq}$ reference frames.}
	\label{Fig:DQdq}
\end{figure}

Instead of selecting a common reference frame rotating at the constant synchronous frequency for all signals under consideration, one may choose time-varying and signal-specific frequencies for the rotating reference frames. In that spirit, consider a rotating reference frame with frequency $\omega(t)$, and the corresponding projection, $X'$:
\begin{equation}\label{eq:dqintro}
    X' = |X'|\mathrm{e}^{\jmath\rho'}.
\end{equation}
The relationship between $X'$ and $\overrightarrow{X}$ is captured by
\begin{align} \label{eq:dqdef}
\begin{split}
    \overrightarrow{X} = X' \mathrm{e}^{\jmath \int_{\tau = 0}^{t} \omega(\tau) \mathrm{d}\tau} = |X'| \mathrm{e}^{\jmath (\int_{\tau = 0}^{t} \omega(\tau) \mathrm{d}\tau + \rho')}.
\end{split}
\end{align}
\noindent The following remarks are in order: 
\begin{itemize}
\renewcommand{\labelitemi}{\tiny$\blacksquare$}
    \item In the above reference-frame representations, given an arbitrary zero-sum $3$-phase signal, the quantities $(\overrightarrow{X},|X|,\rho,|X'|,\rho', \omega)$ are all functions of time, but we mask the time dependence in the notation in the interest of clarity. Given this time dependence, $X$ and $X'$ can be referred to as \textit{dynamic phasors}. {In prior art, dynamic phasors have been referenced in terms of Fourier-series representations of periodic signals with time-varying Fourier-series coefficients~\cite{Sanders, Stankovic}. We have leveraged the definitions in~\cite{VSub,DVR1} and employed a polar-form-based approach for the model development. }
    \item The representation~\eqref{eq:DQintro} is often termed as the global $\mathrm{DQ}$ reference frame. The term \textit{global} stems from the fact that the projection involves a singular frequency. The choice of this frequency is motivated by the underlying observation that networks of different types and scales (from bulk grids to microgrids) operate (by design or by control) close to $\omega_\mrs$. 
    \item The representation~\eqref{eq:dqintro} is often termed as the local $\mathrm{dq}$ reference frame. The term \emph{local} stems from the fact that such representations are typically used for expressing resource-specific quantities across a network. Specifically, consider $N$ resources hosted at nodes of a power network. The control design for individual resources is typically done in stand-alone fashion based on local measurements. Thus, these resources assume their own individual reference-frame frequencies, stacked in $\bomega\in\mathds{R}^{N\times 1}$, to express the related voltage and current signals. In fact, it is a common practice for $\bomega$ to be determined as the estimates of local node-voltage frequencies (imposed or assumed).
    \item Comparing~\eqref{eq:DQdef} and~\eqref{eq:dqdef}, we can infer
\begin{subequations}\label{eq:compare}
\begin{align}
    |X| &= |X'|, \\
    \omega_\mathrm{s}  t + \rho &= \int_{\tau = 0}^{t} \omega(\tau) \mathrm{d}\tau + \rho'.\label{seq:compareB}
\end{align}
\end{subequations}
\end{itemize}
Figure~\ref{Fig:DQdq} illustrates the reference-frame representations described above.
\subsubsection{Sinusoidal Steady-state Operation \& Phasors}
Next, we examine the aforementioned reference frames for representing sinusoidal steady-state signals. Particularly, in steady state, we assume the $3$-phase signals take the form: 
\begin{align}\label{eq:abcSS}
    \bx=\overline{x}\begin{bmatrix}\cos(\omega_\mathrm{ss}t+\overline{\rho})\\
    \cos(\omega_\mathrm{ss}t+\overline{\rho}-\frac{2\pi}{3})\\
    \cos(\omega_\mathrm{ss}t+\overline{\rho}-\frac{4\pi}{3})\end{bmatrix},
\end{align}
where, $\omega_\mathrm{ss}$ is a common constant steady-state frequency (not necessarily equal to $\omega_\mathrm{s}$), and $\overline{x}$ is the constant amplitude. Applying the space-phasor definition~\eqref{eq:SPdef} to the sinusoidal steady-state signal provides
\begin{equation}\label{eq:XphasorSS}
    \overrightarrow{X}=\frac{3}{2}\overline{x}\mathrm{e}^{\jmath(\omega_\mathrm{ss}t+\overline{\rho})}.
\end{equation}
The corresponding global $\mathrm{DQ}$ representation from~\eqref{eq:DQintro} is
$$X=\frac{3}{2}\overline{x}\mathrm{e}^{\jmath((\omega_\mathrm{ss}-\omega_\mrs)t+\overline{\rho})},$$ which establishes constant magnitude and a linear time-variation in phase,
\begin{subequations}\label{eq:DQss}
\begin{align}
    |X| &=\frac{3}{2}\overline{x}, \\
    \rho &=(\omega_\mathrm{ss}-\omega_\mrs)t+\overline{\rho}.
\end{align}
\end{subequations}
The local $\mathrm{dq}$ representation from~\eqref{eq:dqintro} is
$$X'=\frac{3}{2}\overline{x}\mathrm{e}^{\jmath(\omega_\mathrm{ss}t+\overline{\rho})} \mathrm{e}^{-\jmath \int_{\tau = 0}^{t} \omega(\tau) \mathrm{d}\tau}.$$
The magnitude of $X'$ is given by
\begin{subequations} \label{eq:dqss}
\begin{equation}\label{eq:dqMagSS}
|X'|=\frac{3}{2}\overline{x}.
\end{equation}
To quantify the phase $\rho'$, it turns out that we will need to track the time to reach sinusoidal steady state, a term we denote by $t_\mathrm{ss}$. In particular,  $\omega(\tau)=\omega_\mathrm{ss}$ for $\tau\geq t_\mathrm{ss}$. Hence, starting from~\eqref{seq:compareB}, we obtain
\begin{align}\label{eq:dqAngSS}
    \rho'&=\omega_\mathrm{s}t + \rho-\int_{\tau = 0}^{t} \omega(\tau) \mathrm{d}\tau \nonumber \\
    &=\omega_\mathrm{s}t + (\omega_\mathrm{ss}-\omega_\mrs)t+\overline{\rho}-\int_{\tau = 0}^{t} \omega(\tau) \mathrm{d}\tau \nonumber \\
    &= \omega_\mathrm{ss}t+\overline{\rho}-\int_{\tau = 0}^{t_\mathrm{ss}} \omega(\tau) \mathrm{d}\tau -\omega_\mathrm{ss}(t-t_\mathrm{ss}) \nonumber \\
    &=\overline{\rho}-c(t_{\mathrm{ss}}),
\end{align}
\end{subequations}
where, the second equality follows from the steady-state condition on global $\mathrm{DQ}$ phase $\rho$; the third equality partitions the range of integration to pre- and post-steady-state; and finally, the fourth equality identifies and collects the constant terms as
\begin{align*}
    c(t_{\mathrm{ss}}) = \int_{\tau = 0}^{t_\mathrm{ss}} \omega(\tau) \mathrm{d}\tau -\omega_\mathrm{ss}t_\mathrm{ss}.
\end{align*}
These steps establish that the local $\mathrm{dq}$ phase, $\rho'$, assumes a constant value in steady state; the constant depends on the time to reach steady state, $t_\mathrm{ss}$, the steady-state frequency, $\omega_\mathrm{ss}$, and the phase offset, $\overline{\rho}$, in the $\mathrm{abc}$ representation~\eqref{eq:abcSS}. 

For steady-state analysis of a three-phase power system, it is customary to invoke \emph{phasor} representation, which involves constant, complex quantities. The phasor representation for the signal $\bx$ in~\eqref{eq:abcSS} is given by:
\begin{equation}\label{eq:PhasorDef}
    \overline{X} = \frac{\overline{x}}{\sqrt{2}} \mathrm{e}^{\jmath \overline{\rho}},
\end{equation}
where the factor of $\sqrt{2}$ aligns the phasor magnitude with the root mean square (RMS) of the per-phase signals in~\eqref{eq:abcSS}. With the aid of~\eqref{eq:PhasorDef} and \eqref{eq:DQss}, we see that $\overline{X}$ and $X$ are related via
\begin{equation}\label{eq:PhasorDQ}
    \overline{X} = \frac{\sqrt{2}}{3} X \mathrm{e}^{\jmath(\omega_\mathrm{s} - \omega_\mathrm{ss})t}.
\end{equation}
Similarly, with the aid of~\eqref{eq:PhasorDef} and~\eqref{eq:dqss}, we see that $\overline{X}$ and $X'$ are related via
\begin{equation}
    \overline{X} = \frac{\sqrt{2}}{3} X' \mathrm{e}^{\jmath c(t_\mathrm{ss})}.
\end{equation}

\subsection{Defining Power in Different Reference Frames}
Consider three-phase voltages and currents: $\bv = [v_\mathrm{a}, v_\mathrm{b}, v_\mathrm{c}]^\top$ and $\bi = [i_\mathrm{a}, i_\mathrm{b}, i_\mathrm{c}]^\top$. The instantaneous power, $P$, for this signal set is given by
\begin{equation} \label{eq:p-abc}
    P = \bv^\top \bi.
\end{equation}
 From~\eqref{eq:ABC-phasor}, it follows that
\begin{equation*}
    \bv =(\sfrac{2}{3}) \bGamma^\top [\Re(\overrightarrow{V}) \Im(\overrightarrow{V})]^\top, \quad \bi = (\sfrac{2}{3})\bGamma^\top [\Re(\overrightarrow{I}) \Im(\overrightarrow{I})]^\top,
\end{equation*}
which when substituted in~\eqref{eq:p-abc} provides the alternate formulation 
\begin{equation} \label{eq:p-space-phasor}
    P = \frac{2}{3} \left(\Re(\overrightarrow{V}) \Re(\overrightarrow{I}) + \Im(\overrightarrow{V})\Im(\overrightarrow{I})\right).
\end{equation}
Note that we can express the above compactly as: 
\begin{equation} \label{eq:p-space-phasor-compact}
    P = \frac{2}{3} \left(\Re(\overrightarrow{V} \cdot \overrightarrow{I}^*)\right).
\end{equation}
From a first-principles point of view, the product of time-domain voltages and currents is simply instantaneous power. However, in the interest of preserving symmetry with definitions of reactive power (to follow), we will refer to this as instantaneous active power.
The above definition of instantaneous active power \emph{enticingly invites} the following definition of instantaneous reactive power:
\begin{equation} \label{eq:q-space-phasor-compact}
    Q = \frac{2}{3} \left(\Im(\overrightarrow{V} \cdot \overrightarrow{I}^*)\right).
\end{equation}
Interestingly, leveraging~\eqref{eq:phasor-ABC}, we can work out the following dual to~\eqref{eq:p-abc} wherein $Q$ is expressed in terms of $\mathrm{abc}$ quantities
\begin{equation} \label{eq:q-abc}
    Q = \bv^\top \bGamma^{\top}\boldsymbol{\Upsilon}\bGamma \bi.
\end{equation}
With an eye towards completeness, one can define the instantaneous complex-apparent power: 
\begin{equation} \label{eq:s-abc}
S = P + \jmath Q = \frac{2}{3} \overrightarrow{V} \cdot \overrightarrow{I}^* = \bv^\top (\mathbbm{I}_3 + \jmath \bGamma^{\top}\boldsymbol{\Upsilon}\bGamma) \bi. 
\end{equation}

Let us now move on to the representation of the power-related quantities in direct-quadrature reference frames. Corresponding to the three-phase voltage and current waveforms, $\bv$ and $\bi$, denote the direct-quadrature reference-frame transformations by: $V = |V|\mathrm{e}^{\jmath\rho_v}, V'=|V'|\mathrm{e}^{\jmath\rho_v'}$ and $I = |I|\mathrm{e}^{\jmath\rho_i}, I'=|I'|\mathrm{e}^{\jmath\rho_i'}$, respectively. Substituting the relationships in~\eqref{eq:DQdef} and~\eqref{eq:dqdef} in~\eqref{eq:p-space-phasor-compact}--\eqref{eq:q-space-phasor-compact} establishes the following identities:
\begin{subequations}
\begin{align} 
    P &= (\sfrac{2}{3})\Re(V I^*) = (\sfrac{2}{3})|V||I|\cos(\rho_v - \rho_i) \nonumber  \\
    &= (\sfrac{2}{3})\Re(V' (I')^*) =(\sfrac{2}{3}) |V'||I'|\cos(\rho_v' - \rho_i'), \label{eq:P-DQdq} \\
    Q &= (\sfrac{2}{3})\Im(V I^*) =(\sfrac{2}{3}) |V||I|\sin(\rho_v - \rho_i)  \nonumber \\
    &=(\sfrac{2}{3}) \Im(V' (I')^*) = (\sfrac{2}{3})|V'||I'|\sin(\rho_v' - \rho_i'), \label{eq:Q-DQdq}\\
    S &=(\sfrac{2}{3}) V I^* =(\sfrac{2}{3}) V' (I')^*. \label{eq:S-DQdq}
\end{align}
\end{subequations}
Finally, consider sinusoidal steady-state operation with the voltages and currents assuming the form: 
\begin{align*}
    \bv=\overline{v}\begin{bmatrix}\cos(\omega_\mathrm{ss}t+\overline{\rho}_v)\\
    \cos(\omega_\mathrm{ss}t+\overline{\rho}_v-\frac{2\pi}{3})\\
    \cos(\omega_\mathrm{ss}t+\overline{\rho}_v-\frac{4\pi}{3})\end{bmatrix}, \bi=\overline{i}\begin{bmatrix}\cos(\omega_\mathrm{ss}t+\overline{\rho}_i)\\
    \cos(\omega_\mathrm{ss}t+\overline{\rho}_i-\frac{2\pi}{3})\\
    \cos(\omega_\mathrm{ss}t+\overline{\rho}_i-\frac{4\pi}{3})\end{bmatrix}.
\end{align*}
The phasor forms of the voltage and current are given by: $\overline{V} = \frac{\overline{v}}{\sqrt{2}} \mathrm{e}^{\jmath \overline{\rho}_v}$ and $\overline{I} = \frac{\overline{i}}{\sqrt{2}} \mathrm{e}^{\jmath \overline{\rho}_i}$. In this regime, the following identities can be straightforwardly derived:
\begin{subequations}
\begin{align} 
    P &= 3\Re(\overline{V} \cdot \overline{I}^*) = \frac{3}{2}\overline v \overline i \cos(\overline{\rho}_v - \overline{\rho}_i), \label{eq:P-phasor} \\
    Q &= 3 \Im(\overline{V} \cdot \overline{I}^*) = \frac{3}{2}\overline v \overline i \sin(\overline{\rho}_v - \overline{\rho}_i),  \label{eq:Q-phasor}\\
    S &= 3 \overline{V} \cdot \overline{I}^*. \label{eq:S-phasor}
\end{align} 
\end{subequations}
The following remarks are in order: 
\begin{itemize}
\renewcommand{\labelitemi}{\tiny$\blacksquare$}
    \item Instantaneous active and reactive power (and consequently complex apparent power) are invariant to the representation in the global $\mathrm{DQ}$ or local $\mathrm{dq}$ reference frames.
    \item In all reference frames, we note that $P, Q, S$ do not depend on time under sinusoidal stead-state. This is a widely recognized attribute of $3$-phase systems but is not true in general (e.g., in single-phase systems or in a transient sense).
    \item While instantaneous (active) power is a principled quantity related to the potential to do useful work, instantaneous reactive power and instantaneous complex-apparent power are clearly offshoots and mathematical crutches. Undeniably, the averaged versions of these have a home in power engineering parlance, pedagogy, and practice, and several interpretations have been offered (some bordering on amusing~\cite{WinNT}). Instantaneous versions, on the other hand, are not as well studied. (Exceptions include~\cite{Peng_PQ, Dai_PQ}.) 
\end{itemize}
\begin{figure*}
			\centering
			{\includegraphics[width = 1\textwidth]{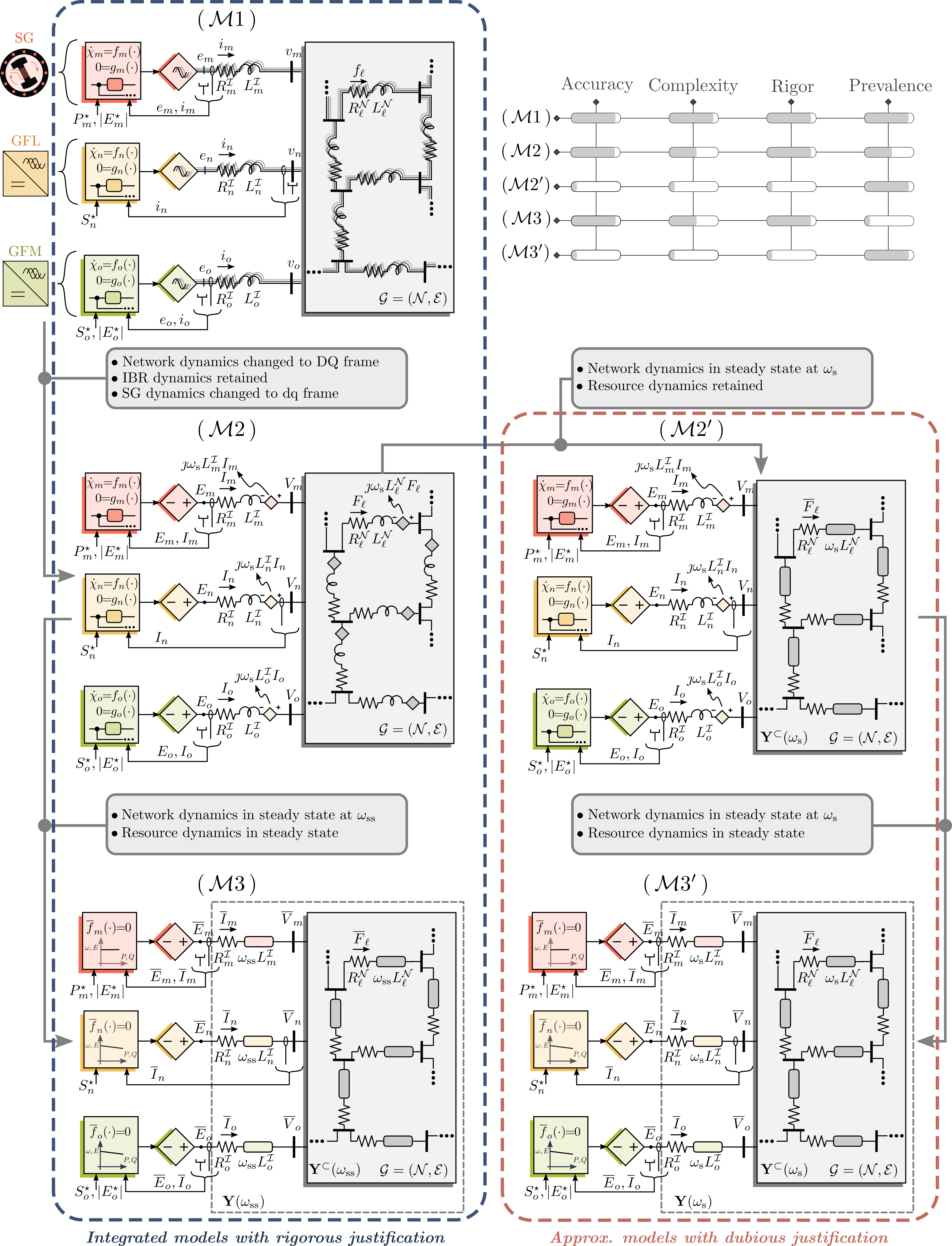}} 
	        \caption{Suite of integrated system models.}
	\label{Fig:Models}
\end{figure*}


\section{Integrated System Models}\label{sec:GenModels}
In this section, we set the stage to present the different types of network and resource models. This includes a high-level summary of all integrated system models alongside an overview of the network models with abstract resource representations to appreciate the subsequent developments. 

\subsection{Summary}
The full suite of integrated system models relevant to the setting of interconnected networks of generators and IBRs is graphically illustrated in Fig.~\ref{Fig:Models}. A thematic connection across all models is that the resources are represented as controlled voltage sources, yielding a cohesive macro-level description. As a precursor, we provide a brief overview of the five classes of models in Fig.~\ref{Fig:Models} below:
\begin{itemize}
\renewcommand{\labelitemi}{\tiny$\blacksquare$}
    \item $(\mathcal{M}1)$: This model is a lumped-parameter $\mathrm{abc}$ description of network and resources. All signals are represented as time-domain real-valued waveforms in the $\mathrm{a}$, $\mathrm{b}$, and $\mathrm{c}$ phases. IBR models are in local $\mathrm{dq}$ reference frames due to the involved controllers; $\mathrm{dq}$-$\mathrm{abc}$ reference-frame transformations interface these to the three-phase network description. In the literature,~$(\mathcal{M}1)$ is commonly referred to as an electromagnetic transient~(EMT) model. We emphasize that our intention is to ensure this model is presented in a true-to-form fashion. By this, we mean that modeling approximations and representation of dynamics in reference frames are kept to a minimum unless absolutely necessary or inherently present by virtue of controller implementation.
    \item $(\mathcal{M}2)$: Recognizing the redundancy in representing all three phases and the computational burden incurred in bookkeeping sinusoidally varying signals, direct-quadrature rotating reference-frame transformations are applied across the board to yield~$(\mathcal{M}2)$. The network dynamics are represented in a global $\mathrm{DQ}$ reference frame at electrical radian synchronous frequency, $\omega_\mathrm{s}$, while the resources are modeled in individual local $\mathrm{dq}$ reference frames as in~$(\mathcal{M}1)$. In this case, $\mathrm{dq}$-$\mathrm{DQ}$ reference-frame transformations interface all resources with the network. For balanced settings, with identical initialization, the results obtained via simulating $(\mathcal{M}1)$ coincide with that of $(\mathcal{M}2)$. 
    \item $(\mathcal{M}3)$: This is a phasor model that is applicable to sinusoidal steady state. Neither $(\mathcal{M}1)$ nor $(\mathcal{M}2)$ presume sinusoidal excitation, however, this assumption is central to the derivation of $(\mathcal{M}3)$, and all involved signals are phasors in the vein of~\eqref{eq:abcSS}. Notably, the sinusoidal steady state that underscores the formulation of $(\mathcal{M}3)$ need not be (in fact, it is unlikely to ever uniformly be) at the electrical radian synchronous frequency, $\omega_\mathrm{s}$. A nettling analytical inconvenience that results is that network impedance depend on steady-state frequency, which in itself, is a function of how resources are actuated. References to power-flow models and/or phasor models in the literature ought to be in the context of $(\mathcal{M}3)$, but, these actually refer to an approximate version, $(\mathcal{M}3')$, that dispels the inconvenience discussed above. 
    \item $(\mathcal{M}2')$ \& $(\mathcal{M}3')$: These are widely referenced approximate models with murky origin stories. In particular, $(\mathcal{M}2')$ retains dynamic models for resources, but curiously, models the network in sinusoidal steady state at $\omega_\mathrm{s}$ so that one can leverage the computational ease offered by phasors. The oft-referenced differential algebraic equation (DAE) models for power-system dynamics are in this form. $(\mathcal{M}2')$ is typically justified with time-scale separation arguments in bulk grids with predominantly synchronous generators. However, one loses the luxury of crisp separation of timescales in complex networks with a variety of resource types. (This has been recognized in the literature in various contexts~\cite{Ajala2018,ajala2018phdthesis,Vorobev-2018,Gross-2019,Rodrigo-2020,Ajala-2021}.) Finally, $(\mathcal{M}3')$ models the resources in steady state while retaining the $\omega_\mathrm{s}$ steady-state model for the network. $(\mathcal{M}3')$ inevitably inherits the accuracy limits of $(\mathcal{M}2')$ in cases where the steady-state frequency of the network may not be $\omega_\mathrm{s}$. The ubiquitous power-flow equations apply to the setting of~$(\mathcal{M}3')$. (See Section~\ref{sec:Approx} for details on~$(\mathcal{M}2')$ and~$(\mathcal{M}3')$.)
\end{itemize}
The table tucked away in the top-right corner of Fig.~\ref{Fig:Models} qualitatively compares the five models discussed above on the grounds of: how accurately they capture dynamics (at the associated timescale); what level of computational complexity is involved in implementation and simulation; whether the formulations are rigorous and can be justified from a first-principles point of view; and finally, how pervasive they are in power-system parlance, pedagogy, and practice. Illustrations of these attributes in the table are to be interpreted in a subjective and comparative sense (particularly in the case of the approximate models $(\mathcal{M}2'), (\mathcal{M}3')$).

At this early juncture, it is worth emphasizing a cautionary note pertaining to the circuit representation of generators and IBRs in steady state. The steady-state models in $(\mathcal{M}3),(\mathcal{M}3')$ include references to the steady-state frequency, $\omega_\mathrm{ss}$. (This is explicitly obvious in $(\mathcal{M}3)$; while in $(\mathcal{M}3')$, we will see this dependence arises from resource terminal behavior in steady state.) This quantity is not a constant (in particular, it is not equal to $\omega_\mathrm{s}$ given that the examined setting does not include secondary control that continually aspires to restore synchronous operation); rather, it is a function of network operation, and therefore, has to be solved for extrinsically. (See, e.g., Section~\ref{sec:PF} and Section~\ref{sec:omegass}.) As such, the models~$(\mathcal{M}3),(\mathcal{M}3')$ are not self contained and cannot be directly implemented and solved in a circuit simulator.  

\subsection{Overview of Network-resource System}
We consider a three-phase balanced $RL$ network with $N$ nodes and $E$ edges. Model the network as a directed graph $\mcG=(\mcN, \mcE)$ with node-set $\mcN=\{1,\dots,N\}$ and edge-set $\mcE$. To capture the topology of the power network modeled as graph $\mcG$, we begin with arbitrarily assigning the directions for edges $e\in\mcE$. The connectivity of the graph $\mcG$ can then be conveyed using the incidence matrix $\bB\in\{0,\pm 1\}^{N\times E}$ with entries defined as
\begin{equation*}
{B}_{k,e}:=
\begin{cases}
+1&,~k=m\\
-1&,~k=n\\
0&,~\text{otherwise}
\end{cases}~\forall~e=(m,n)\in\mcE.
\end{equation*} 
We assume without loss of generality that all network nodes in $\mcN$ host exactly one generator or IBR. Construct $E\times E$ diagonal matrices $(\bR^\mcN, \bL^\mcN)$ with the diagonal entries representing the line resistances and inductances, respectively. Similarly, $N\times N$ diagonal matrices $(\bR^\mcI, \bL^\mcI)$ collect the resistances and inductances that interface the generators and IBRs to the network. For IBRs, these could subsume output-filter values. 

To preserve notational simplicity, we refrain from including shunt capacitive elements in the line models, complex filter arrangements (e.g., $LCL$ filters) in the inverters, and transformers in the network. Note also that the ratings of synchronous generators are generally orders-of-magnitude higher than those of individual inverters. Typically, scores of inverters are organized in plants to scale capacity. We refrain from discussing aggregate models, but point to~\cite{Purba-2017,Khan-2018,Vijayshankar-2019,Purba-2019,Purba-2020} (and references therein) for pertinent effort in this direction. Finally, we note that all parameters and signals---network and resource related---are expressed in SI units unless stated explicitly.


The governing dynamics for a resource connected to bus $n\in\mcN$ can be expressed in general form as
\begin{subequations}\label{eq:IBRdyn}
\begin{align}
    \diff{\bchi_n}{t}&=\bof_n(\bchi_n, \by_n),\\
    \bzero&=\bg_n(\bchi_n,\by_n),
\end{align}
\end{subequations}
where vectors $\bchi_n, \by_n$ collect internal dynamic and algebraic states for the resources. The terminal voltage $\be$ corresponding to $\mathrm{abc}$ representation, or the tuple $(|E'|,\delta',\omega)$ corresponding to $\mathrm{dq}$ representation may feature in either $\bchi_n$ or $\by_n$, based on resource type (SG, GFL IBR, GFM IBR) and integrated system model $(\mathcal{M}1)$-$(\mathcal{M}2)$. We clarify this in forthcoming discussions.

Next, we consider the setting where the resource and network dynamics evolve to a steady state. We do not comment on the conditions that can guarantee the existence of a frequency-synchronized equilibrium, but point to the fact that this has been studied for models of varying complexity in the literature~\cite{Tim08droop,Dorfler16hierarchy,Xin11PVcontrol,Ajala-CDC,hodge2022synch}. However, analytical results are lacking for the complex interconnected setting and heterogeneous mix of resources we examine in this work; this is an open problem and deserves scrutiny. Assuming that a steady-state operating point exists for the networked collection, frequency synchronized operation in sinusoidal steady state corresponds to assuming $\bomega=\omega_{\mathrm{ss}}\bone$, with $\omega_{\mathrm{ss}}$ being the common steady-state frequency of all network and resource variables. In this regime, the terminal behavior of a specific resource at node $n$ can be modeled using the algebraic form
\begin{equation}\label{eq:IBRss}
    \overline{\bof}_n(\overline{E}_n,|\overline{E}_n^{\star}|, \overline{I}_n,P^\star_n,Q^\star_n,\omega_\mathrm{ss})=\bzero,
\end{equation}
where, the function $\overline{\bof}_n$ is parametrized by a set of known parameters and reference setpoints. (Details are in Section~\ref{sec:Resource}.)

In what follows, we discuss the network and resource models that complete the picture in the specification of $(\mathcal{M}1)$--$(\mathcal{M}3)$ \& $(\mathcal{M}2')$--$(\mathcal{M}3')$. First, we present the network models  in the same order as the presentation of reference frames in the preliminaries, i.e., starting with the $\mathrm{abc}$ model $(\mathcal{M}1)$ in Section~\ref{sec:abc}, moving on to the direct-quadrature model $(\mathcal{M}2)$ in Section~\ref{sec:DQ}, and finally on to the sinusoidal steady-state model $(\mathcal{M}3)$ in Section~\ref{sec:phasor}. Next, we present the resource models organized by resource type; in particular, we cover the synchronous generator model in Section~\ref{sec:SG}, GFL IBR model in Section~\ref{sec:GFL}, and GFM IBR model in Section~\ref{sec:GFM}. In each case, we strive to present the resource models in terms of how they fit in with the network models, i.e., $\mathrm{abc}$, direct-quadrature, and steady state. We reserve our discussion of approximate models $(\mathcal{M}2')$--$(\mathcal{M}3')$ to Section~\ref{sec:Approx}. 

\section{Network Models}\label{sec:Network}
{For network modeling, we consider short transmission lines (comprising of $RL$ impedances and devoid of shunt capacitances) for simplicity. (However, the models and modeling approach presented here are generic; the ideas expressed and inferences derived later are readily extendable to other settings, e.g., with medium transmission-line models that include shunt line capacitances.)} In what follows, we spell out the governing equations for the network $RL$ lines and resource-interfacing $RL$ branches in $\mathrm{abc}$ and $\mathrm{DQ}$ reference frames under dynamic conditions; and follow that up with a discussion on algebraic equation models that emerge in sinusoidal steady state. With each model, we clearly illustrate the connection to three-phase signals, $\be$, tuple $(|\bE'|, \bdelta', \bomega)$ relevant to direct-quadrature reference-frame modeling, and phasor form, $\overline{\bE}$, based on resource type, integrated system model, and timescale.
\begin{figure*}[]
	\centering
	{\includegraphics[width = 1\linewidth]{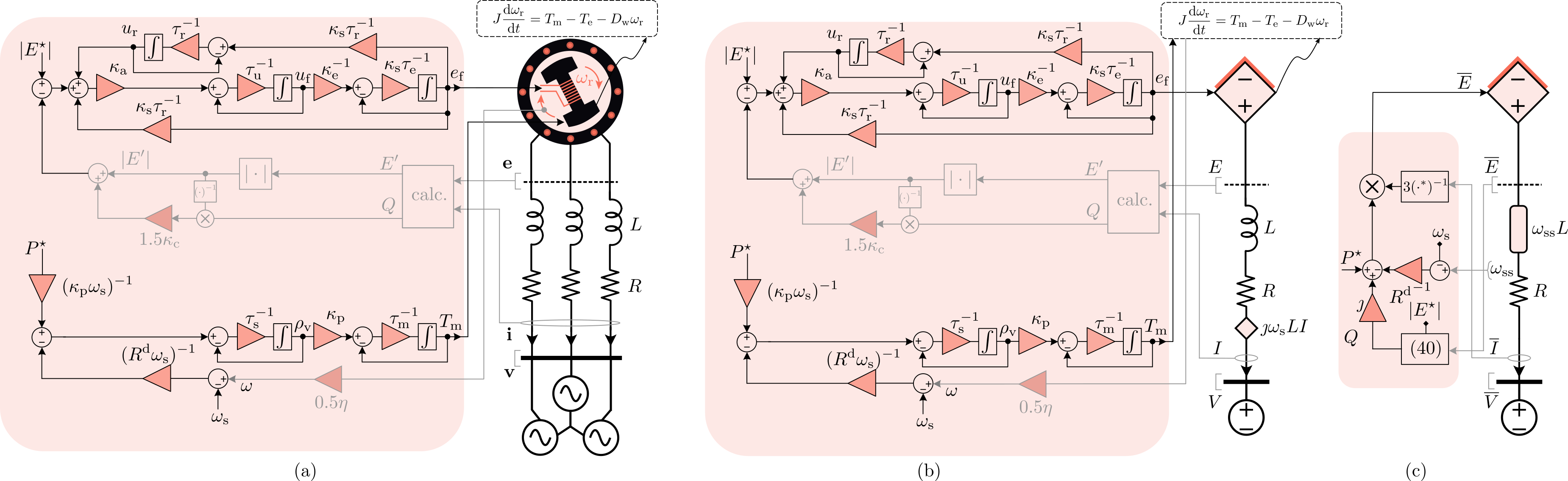}} 
    \caption{Synchronous-generator models: (a)~$\mathrm{abc}$ reference-frame model compatible with~$(\mathcal{M}1)$, (b)~direct-quadrature reference-frame model compatible with~$(\mathcal{M}2)$-$(\mathcal{M}2')$, (c)~sinusoidal steady-state phasor model compatible with~$(\mathcal{M}3)$-$(\mathcal{M}3')$. Grayed-out portions are part of auxiliary control loops and not critical to appreciating central ideas.}
	\label{Fig:SG}
\end{figure*} 
\begin{figure*}[]
	\centering
	{\includegraphics[width = 1\linewidth]{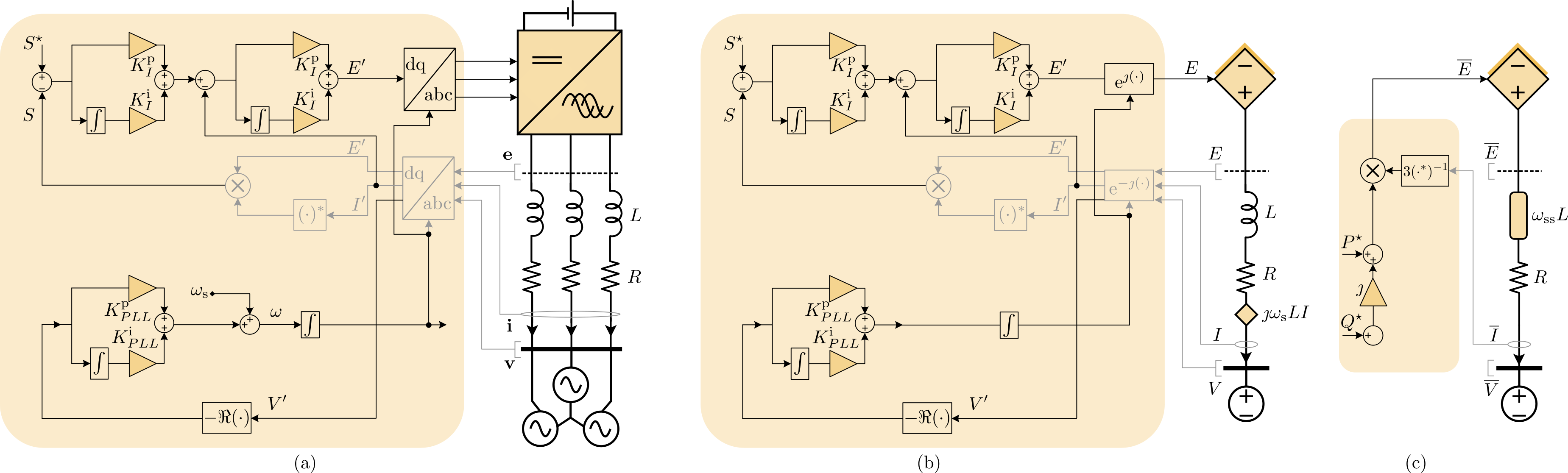}} 
    \caption{Grid-following (GFL) IBR models: (a)~$\mathrm{abc}$ reference-frame model compatible with~$(\mathcal{M}1)$, (b)~direct-quadrature reference-frame model compatible with~$(\mathcal{M}2)$-$(\mathcal{M}2')$, (c)~sinusoidal steady-state phasor model compatible with~$(\mathcal{M}3)$-$(\mathcal{M}3')$. Grayed-out portions are part of auxiliary control loops and not critical to appreciating central ideas.}
	\label{Fig:GFL}
\end{figure*} 
\begin{figure*}[]
	\centering
	{\includegraphics[width = 1\linewidth]{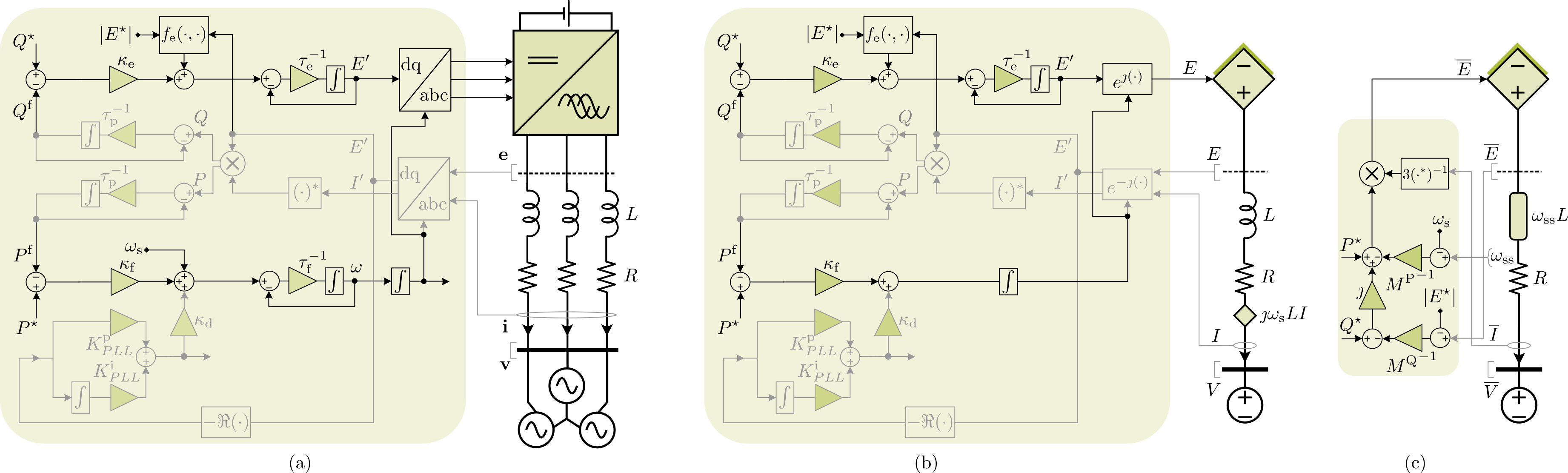}} 
    \caption{Grid-forming (GFM) IBR models: (a)~$\mathrm{abc}$ reference-frame model compatible with~$(\mathcal{M}1)$, (b)~direct-quadrature reference-frame model compatible with~$(\mathcal{M}2)$-$(\mathcal{M}2')$, (c)~sinusoidal steady-state phasor model compatible with~$(\mathcal{M}3)$-$(\mathcal{M}3')$. Grayed-out portions are part of auxiliary control loops and not critical to appreciating central ideas.}
	\label{Fig:GFM}
\end{figure*}

\subsection{Three-phase $\mathrm{abc}$ Model~$(\mathcal{M}1)$} \label{sec:abc}
To present the per-phase dynamics of $RL$ lines and interfaces, let us denote the $\mra$-phase network-node voltage vector by $\bv_\mra$, line currents by $\bof_\mra$, resource voltages by $\be_\mra$, and resource current injections to the network by $\bi_\mra$. The dynamics of the voltage-behind-$RL$-line subsystem in $\mathrm{abc}$ coordinates can then be expressed as: 
\begin{align} \label{eq:N1} \tag{$\mathcal{M}1$}
\begin{split}
    \bL^\mcI \diff{\bi_\mathrm{a}}{t} &= \be_\mathrm{a} - \bv_\mathrm{a} -\bR^\mcI \bi_\mathrm{a}, \\
    \bL^\mcN \diff{\bof_\mathrm{a}}{t} &=\bB^\top \bv_\mathrm{a} -\bR^\mcN \bof_\mathrm{a}, \\
    \bi_\mathrm{a} &= \bB \bof_\mathrm{a},
    \end{split}
\end{align}
where, leveraging the definitions in~\eqref{eq:ABC-phasor} and \eqref{eq:dqdef}, the $\mathrm{a}$-phase voltages $\be_\mathrm{a}$ are given by: 
\begin{subequations}
\begin{equation}\label{eq:ea}
\be_\mathrm{a} =\frac{2}{3} |\bE'| \circ \cos(\textstyle \int_{\tau = 0}^{t} \bomega(\tau) \mathrm{d}\tau + \bdelta').
\end{equation}
Dynamics of the $\mathrm{b}$ and $\mathrm{c}$ phase circuits are analogous to~\eqref{eq:N1}, except with excitations: 
\begin{align}
\be_\mathrm{b} &= -\frac{1}{3} |\bE'| \circ \cos(\textstyle \int_{\tau = 0}^{t} \bomega(\tau) \mathrm{d}\tau + \bdelta') \nonumber \\
&\quad +\frac{1}{\sqrt{3}} |\bE'| \circ \sin(\textstyle \int_{\tau = 0}^{t} \bomega(\tau) \mathrm{d}\tau + \bdelta'), \label{eq:eb}\\
\be_\mathrm{c} &= -\frac{1}{3} |\bE'| \circ \cos(\textstyle \int_{\tau = 0}^{t} \bomega(\tau) \mathrm{d}\tau + \bdelta') \nonumber \\
&\quad -\frac{1}{\sqrt{3}} |\bE'| \circ \sin(\textstyle \int_{\tau = 0}^{t} \bomega(\tau) \mathrm{d}\tau + \bdelta').\label{eq:ec}
\end{align} 
\end{subequations}
We will find that for the case of SGs, three-phase signals $\be_\mathrm{a}, \be_\mathrm{b}, \be_\mathrm{c}$ correspond to the stator voltages, the dynamics of which are tied to the underlying electromechanical phenomena and companion controllers (voltage regulators and frequency governors). On the other hand, for IBRs, controls are implemented in local $\mathrm{dq}$ reference frames. The result is that the combination of the set $(|\bE'|,\bdelta',\bomega(t))$, and the expressions~\eqref{eq:ea}--\eqref{eq:ec} capture how the inverter terminals in the network are actuated.

\subsection{Direct-quadrature Reference-frame Model~$(\mathcal{M}2)$} \label{sec:DQ}
Adopting $\mathrm{DQ}$ representation, let the resource voltage and currents be denoted as $\bE$ and $\bI$; and network-node voltages and line currents be represented as $\bV$ and $\bF$. Then, the $\mathrm{DQ}$-frame $RL$ dynamics can be expressed as
\begin{align} \label{eq:DQ} \tag{$\mathcal{M}2$}
\begin{split} 
    \bL^\mcI \diff{\bI}{t} &= \bE {-} \bV -\bR^\mcI\bI- \jmath \omega_\mathrm{s} \bL^\mcI \bI,   \\
     \bL^\mcN \diff{\bF}{t} &= \bB^\top\bV -\bR^\mcN \bF -\jmath\omega_\mrs\bL^\mcN\bI,\\
     \bI & =\bB\bF,
\end{split}
\end{align}
where, the $\mathrm{DQ}$ signal, $\bE$, can be expressed using~\eqref{eq:compare} as: 
\begin{equation} \label{eq:E}
    \bE = |\bE'| \circ \mathrm{e}^{\jmath(\int_{\tau = 0}^{t} (\bomega(\tau) - \omega_\mathrm{s} \bone) \mathrm{d} \tau + \bdelta')}.
\end{equation}
In this model, all resource dynamics are deliberately expressed in local $\mathrm{dq}$ reference frames. Recall that this was true for inverters even in $(\mathcal{M}1)$ due to control implementation. For the case of generators, we will see that resorting to a local $\mathrm{dq}$ reference-frame representation significantly simplifies the model complexity. (Details in Section~\ref{sec:SG}.)

\subsection{Sinusoidal Steady-state Phasor Model~$(\mathcal{M}3)$} \label{sec:phasor}
 As premised in introducing~\eqref{eq:IBRss}, under steady state, the resources are modeled as generating complex voltage phasors $\overline{\bE}$ at frequency $\omega_\mrss$. To obtain the steady-state model for $RL$ lines and interfaces, in addition to resource voltage phasor $\overline{\bE}$, we introduce the node voltage, current injection, and line current phasors $(\overline{\bV},\overline{\bI},\overline{\bF})$. From \eqref{eq:PhasorDQ}, we note that the phasors $(\overline{\bV},\overline{\bE},\overline{\bI},\overline{\bF})$ can be obtained from the respective $\mathrm{DQ}$-frame quantities $(\bV, \bE, \bI,\bF)$ by multiplying with $(\sfrac{\sqrt{2}}{3})\mre^{\jmath(\omega_{\mathrm{ss}}-\omega_\mrs)t}$. 
Substituting the steady-state conditions~\eqref{eq:DQss} in \eqref{eq:DQ} and invoking the phasor definitions allow writing the steady-state $RL$ dynamics as algebraic relations:
\begin{align} \label{eq:RLss} \tag{$\mathcal{M}3$}
\begin{split} 
    \overline{\bE} {-} \overline{\bV} &=\left(\bR^\mcI+\jmath\omega_{\mathrm{ss}}\bL^\mcI\right)\overline{\bI},\\
    \bB^\top\overline{\bV} &=\left(\bR^\mcN+\jmath\omega_{\mathrm{ss}}\bL^\mcN\right)\overline{\bF},\\
    \overline{\bI}&=\bB\overline{\bF},
\end{split}
\end{align}
where 
\begin{equation} \label{eq:Ephasor}
   \overline{\bE}=\frac{1}{3}(\diag(\overline{\bI}^*))^{-1}{\bS}.
\end{equation}
Resource terminal behavior---as captured abstractly in~\eqref{eq:IBRss} and to be formalized in Section~\ref{sec:Resource}---appropriately populates both sides of~\eqref{eq:Ephasor}. 

The algebraic model for steady-state conditions gives way to further simplifications. For instance, oftentimes, the focus of steady-state analyses is on the terminal quantities $(\overline{\bE},\overline{\bI})$ and one may seek to simplify \eqref{eq:RLss} to eliminate $(\overline{\bV},\overline{\bF})$. Towards such simplification, let us substitute $\overline{\bV}$ from the first equation in \eqref{eq:RLss} into the second, and solve for $\overline{\bF}$ to obtain
\begin{equation}\label{eq:Fbar}
\overline{\bF}=(\bR^\mcN+\jmath\omega_\mathrm{ss}\bL^\mcN)^{-1}\bB^\top\big(\overline{\bE} {-} \left(\bR^\mcI+\jmath\omega_{\mathrm{ss}}\bL^\mcI\right)\overline{\bI}\big).\end{equation}
Multiplying both sides by $\bB$, and using the last equation in \eqref{eq:RLss}, we get
\begin{equation*}\label{eq:IYV1}
\overline{\bI}=\bB(\bR^\mcN+\jmath\omega_\mathrm{ss}\bL^\mcN)^{-1}\bB^\top\big(\overline{\bE} {-} \left(\bR^\mcI+\jmath\omega_{\mathrm{ss}}\bL^\mcI\right)\overline{\bI}\big),
\end{equation*}
where, the matrix
\begin{equation*}
    \bY^{\subset}(\omega_\mathrm{ss})=\bB(\bR^\mcN+\jmath\omega_\mathrm{ss}\bL^\mcN)^{-1}\bB^\top
\end{equation*}
is the admittance matrix of the network $\mcG$. Solving further for current $\overline{\bI}$ yields
\begin{equation}\label{eq:IYV}
    \overline{\bI}=[\mathbbm{I}_N+\bY^{\subset}(\omega_\mathrm{ss})(\bR^\mcI+\jmath\omega_\mathrm{ss}\bL^\mcI)]^{-1}\bY^{\subset}(\omega_\mathrm{ss})\overline{\bE}.
\end{equation}
Equation~\eqref{eq:IYV} relating $\overline{\bI}$ and $\overline{\bE}$ is the reduced system description (capturing \eqref{eq:RLss}). The matrix
\begin{equation}\label{eq:IYV-def}
    \bY(\omega_\mathrm{ss})=[\mathbbm{I}_N+\bY^{\subset}(\omega_\mathrm{ss})(\bR^\mcI+\jmath\omega_\mathrm{ss}\bL^\mcI)]^{-1}\bY^{\subset}(\omega_\mathrm{ss})
\end{equation}
serves as the equivalent admittance matrix for the entire network including resource $RL$ interconnections. One could alternatively arrive at~\eqref{eq:IYV} by: i)~augmenting the power network $\mcG$ to include the $RL$ interfacing branches; and ii)~invoking Kron reduction to eliminate the network nodes, yielding an equation, equivalent to~\eqref{eq:IYV}, relating $\overline{\bI}$ and $\overline{\bE}$.


\section{Resource Models} \label{sec:Resource}
In this section, we present the resource models for synchronous generators and IBRs. We make a concerted effort to clarify what facets of the presented dynamics arise intrinsically from the form and function of the resource and accompanying controllers, and what facets are to be appreciated as modeling efforts motivated by the desire to seek analytically convenient and/or computationally lean representations. Steady-state models are presented with the intention of capturing network-facing behavior pertinent to~$(\mathcal{M}3)$ and~$(\mathcal{M}3')$. Note that we drop subscripts that index the resource to contain notational complexity in this section. We advocate consuming the content that follows along with Figs.~\ref{Fig:SG},~\ref{Fig:GFL},~\ref{Fig:GFM} which illustrate key aspects of the resource models in terms of their network-facing representation across timescales. (In the illustrations in Fig.~\ref{Fig:SG}--Fig.~\ref{Fig:GFM}, the power stage and primary control loops involving voltage and frequency (or torque in case of SG) are shown in black, while the portions concerning feedback sensing, reference-frame transformation and subsequent computations are shown in gray.) {We have also made a deliberate choice of excluding feedback control loops for certain inner variables in the control systems (particularly those associated with fast timescales) to keep the model order and complexity low while retaining relevant resource dynamics for integrated modeling. We have also not included limiters for inner control variables in the control systems. While they may be always present in practical implementations, they typically come into play only during adverse situations (e.g., inverter overload and faults) and are not invoked during nominal operating conditions. Hence, they have been excluded for the purpose of modeling resources.}

\subsection{Synchronous-generator Model} \label{sec:SG}
The three-phase synchronous generator is composed of a rotor that is equipped with a field winding and several damper windings, a stator that is equipped with three-phase stator windings, a field winding excitation system, a prime mover, and a speed governor.

\subsubsection{Stator \& Rotor Dynamics}
In SI units, let $\bm{\lambda}$ denote flux linkage, let $\be$ denote voltage, let $\bm{i}$ denote current, let $\omega$ denote angular frequency of the rotor in electrical radians per second, let $\theta_\mathrm{r}$ denote angular position of the rotor, in radians, let $e_\mathrm{f}$, $e_1$, $e_2$ denote the external excitation voltages for the field winding and damper windings $1$ and $2$, respectively, and let $T_{\mathrm{m}}$ and $T_{\mathrm{e}}$ denote mechanical and electrical torques, respectively. Then, as detailed in \cite[pp.~20--30]{sauer2017power} and \cite[pp. 145--148, 277]{KrauseWasynczuk2013}, using Kirchoff's, Faraday's, and Newton's laws, we have that the dynamics of a balanced three-phase synchronous generator with two damper windings can be described by
\begin{subequations}\label{eq:windings}
\begin{align}
    \diff{\bm{\lambda}}{t} &= - \bm{i} r + \be,\label{eq:lambda_abc1}\\
    \diff{\lambda_\mathrm{f}}{t} &= - i_\mathrm{f} r_\mathrm{f} + e_\mathrm{f},\\
    \diff{\lambda_{1}}{t} &= - i_{1} r_{1} + e_{1},\\
    \diff{\lambda_{2}}{t} &= - i_{2} r_{2} + e_{2},\\
    \diff{\theta_\mathrm{r}}{t} &= \frac{2}{\eta}\omega,\\
    J\frac{2}{\eta}\diff{\omega}{t} &= T_{\mathrm{m}} - T_{\mathrm{e}} - D_\mathrm{w}\frac{2}{\eta}\omega,\\
    \bm{\lambda} &= (l_\mathrm{ls}\mathbbm{I}_{3}+L_\mathrm{m}(\theta_\mathrm{r})) \bm{i} + l_\mathrm{sf}\boldsymbol{\sin}(\eta\theta_\mathrm{r}) i_\mathrm{f}\nonumber\\
    &+ l_\mathrm{s1}\boldsymbol{\sin}(\eta\theta_\mathrm{r}) i_\mathrm{1} + l_\mathrm{s2}\boldsymbol{\cos}(\eta\theta_\mathrm{r}) i_\mathrm{2},\label{eq:lambda_abc2}\\
    \lambda_\mathrm{f} &= l_\mathrm{sf}\boldsymbol{\sin}(\eta\theta_\mathrm{r})^{\top} \bm{i} + l_\mathrm{ff} i_\mathrm{f} + l_\mathrm{f1} i_\mathrm{1},\label{eq:lambda_f}\\
    \lambda_\mathrm{1} &= l_\mathrm{s1}\boldsymbol{\sin}(\eta\theta_\mathrm{r})^{\top} \bm{i} + l_\mathrm{f1} i_\mathrm{f} + l_\mathrm{11} i_\mathrm{1},\label{eq:lambda_1}\\
    \lambda_\mathrm{2} &= l_\mathrm{s2}\boldsymbol{\cos}(\eta\theta_\mathrm{r})^{\top} \bm{i} + l_\mathrm{22} i_\mathrm{2},\label{eq:lambda_2}
\end{align}
\end{subequations} where, $r$ denotes the per-phase resistance of the stator windings, $r_\mathrm{f}$, $r_1$, and $r_2$ denote resistances of the field winding and two damper windings, respectively, $\eta$ denotes the number of magnetic poles of the rotor, $D_\mathrm{w}$ denotes a damping coefficient associated with windage and friction, $l_\mathrm{ls}$ denotes the leakage inductance of the stator windings, $l_\mathrm{sf}$ denotes the mutual inductance between the stator and the field winding, $l_\mathrm{s1}$ and $l_\mathrm{s2}$ denote the mutual inductance between the stator windings and damper windings $1$ and $2$, respectively, $l_\mathrm{f1}$ denotes the mutual inductance between the field winding and damper winding $1$, and $l_\mathrm{ff}$, $l_\mathrm{11}$, and $l_\mathrm{22}$ denote the self inductances of the field winding and damper windings $1$ and $2$, respectively. The inductance matrix $L_\mathrm{m}(\theta_\mathrm{r})$ is defined as:
\begin{align*}
    L_\mathrm{m}(\theta_\mathrm{r}) &= \left[\begin{matrix}
        l_\mathrm{A}-l_\mathrm{B}{\cos}\eta\theta_\mathrm{r} \\
        -\frac{1}{2}l_\mathrm{A}-l_\mathrm{B}{\cos}\big(\eta\theta_\mathrm{r}-\frac{2\pi}{3}\big) \\
        -\frac{1}{2}l_\mathrm{A}-l_\mathrm{B}{\cos}\big(\eta\theta_\mathrm{r}+\frac{2\pi}{3}\big) 
    \end{matrix}\right.\\
    &\hspace{-0.4in}\left.\begin{matrix}
        -\frac{1}{2}l_\mathrm{A}-l_\mathrm{B}{\cos}\big(\eta\theta_\mathrm{r}-\frac{2\pi}{3}\big) & -\frac{1}{2}l_\mathrm{A}-l_\mathrm{B}{\cos}\big(\eta\theta_\mathrm{r}+\frac{2\pi}{3}\big)\\
        l_\mathrm{A}-l_\mathrm{B}{\cos}\big(\eta\theta_\mathrm{r}+\frac{2\pi}{3}\big) & -\frac{1}{2}l_\mathrm{A}-l_\mathrm{B}{\cos}\eta\theta_\mathrm{r}\\
        -\frac{1}{2}l_\mathrm{A}-l_\mathrm{B}{\cos}\eta\theta_\mathrm{r} & l_\mathrm{A}-l_\mathrm{B}{\cos}\big(\eta\theta_\mathrm{r}-\frac{2\pi}{3}\big)
    \end{matrix}\right]
\end{align*} where $l_\mathrm{A}$ and $l_\mathrm{B}$ denote machine constants, with $l_\mathrm{A}>l_\mathrm{B}$ for salient-pole machines and $l_\mathrm{B}=0$ for round-rotor machines.

\subsubsection{Voltage-regulator Dynamics}

Let $u_\mathrm{f}$ and $u_\mathrm{r}$ denote control input of the excitation system, and the rate feedback variable of the voltage regulator. Then, the dynamics of the field excitation system can be described by
\begin{subequations}\label{eq:exciter}
\begin{align}
    \tau_\mathrm{e}\diff{e_\mathrm{f}}{t} &= -\kappa_\mathrm{e}e_\mathrm{f} + u_\mathrm{f},\\
    \tau_\mathrm{u}\diff{u_\mathrm{f}}{t} &= -u_\mathrm{f} + \kappa_\mathrm{a}u_\mathrm{r} - \frac{\kappa_\mathrm{a}\kappa_\mathrm{s}}{\tau_\mathrm{r}}e_\mathrm{f} \nonumber \\&+ \kappa_\mathrm{a}\Big(|E^\star|-|E'|-\frac{3\kappa_\mathrm{c}}{2|E'|}Q\Big),\label{eq:avr-temp}\\
    \tau_\mathrm{r}\diff{u_\mathrm{r}}{t} &= -u_\mathrm{r} + \frac{\kappa_\mathrm{s}}{\tau_\mathrm{r}}e_\mathrm{f},
\end{align}
\end{subequations} where $|E^\star|$ denotes the reference voltage magnitude, $Q$ is the reactive power output of the machine, $\tau_\mathrm{e}$, $\tau_\mathrm{u}$, and $\tau_\mathrm{r}$ denote time constants of the exciter, the voltage regulator, and the rate feedback systems, respectively, ${\kappa}_{\mathrm{e}}$ denotes the exciter field proportional constant, ${\kappa}_{\mathrm{a}}$ denotes the amplifier gain, ${\kappa}_{\mathrm{s}}$ denotes the stabilizer gain for the rate feedback, ${\kappa}_{\mathrm{c}}$ denotes the load compensation term that realizes voltage droop with reactive power. Note that $|E'|$ is the machine terminal voltage magnitude in the local $\mathrm{dq}$ reference frame rotating at $\omega(t)$. We caution that the terms $|E'|$ and $Q$ are featured in the representation~\eqref{eq:avr-temp} motivated by notational simplicity; in practice, these quantities may be realized in different functional forms (with appropriate signal and parametric dependence) that yield the same numerical value (see~\cite[pp.~10--13]{IEEE_ExcMod}).

\subsubsection{Governor Dynamics}
Let $T_\mathrm{m}$ denote the mechanical torque output of the generator, and let $\rho_{\mathrm{v}}$ denote the valve position of the prime mover. The dynamics of the prime mover and speed governor can be described by 
\begin{subequations}\label{eq:governor}
\begin{align}
    \tau_\mathrm{m}\diff{T_\mathrm{m}}{t} &= -T_\mathrm{m} + \kappa_{\mathrm{p}}\rho_{\mathrm{v}},\\
    \tau_\mathrm{s}\diff{\rho_{\mathrm{v}}}{t} &= -\rho_{\mathrm{v}} + \frac{1}{\kappa_{\mathrm{p}}\omega_{\mathrm{s}}}P^\star - \frac{1}{R^\mathrm{d}\omega_{\mathrm{s}}}({\omega}-{\omega_\mathrm{s}}),
\end{align}
\end{subequations} where $\kappa_{\mathrm{p}}$ denotes a scaling constant that represents the ratio of nominal torque to nominal valve position, $\tau_\mathrm{m}$ and $\tau_\mathrm{s}$ denote time constants for the prime mover and speed governor, respectively, $P^\star$ denotes the active power reference, and $R^\mathrm{d}$ denotes the droop coefficient.

Collectively,~\eqref{eq:windings},~\eqref{eq:exciter}, and~\eqref{eq:governor} constitute the $\mathrm{abc}$ reference-frame representation of generator dynamics and would feature accordingly in~$(\mathcal{M}1)$ with $\be$ serving as the interface signal. With the exception of the representation of stator-voltage amplitude and reactive power in the voltage-regulator dynamics discussed above, this model captures all device physics and controller details without any representational simplifications. Figure~\ref{Fig:SG}(a) provides an abstract representation of the model discussed above in the $\mathrm{abc}$ reference frame. Highlighted dynamics are those of the voltage regulator, frequency governor, and mechanical motion of the rotor. These are key to how the generator interfaces with the electrical network. Significant portions of the model (e.g., stator- and field-winding dynamics) are masked or grayed out for simplicity.

What follows next is a local $\mathrm{dq}$ representation of stator and rotor dynamics to simplify the complexity inherent in the stator- and field-winding dynamics. 

\noindent \subsubsection*{Direct-quadrature Reference-frame Transformation} We simplify the expressions in~\eqref{eq:windings} by transforming them to a local $\mathrm{dq}$ reference frame with angular position $\frac{\eta}{2}\theta_\mathrm{r}$. This is achieved by using the relations~\eqref{eq:ABC-phasor} and~\eqref{eq:dqdef}. Explicitly, we have the following relationship between $\mathrm{abc}$ and $\mathrm{dq}$ representations:
\begin{align}
    \bx &= \frac{2}{3}\bGamma^\top \begin{bmatrix} \Re(X'\mathrm{e}^{\jmath \frac{\eta}{2}\theta_\mathrm{r}}) \\ \Im(X'\mathrm{e}^{\jmath \frac{\eta}{2}\theta_\mathrm{r}}) \end{bmatrix}.
\end{align}
Applying the above,~\eqref{eq:lambda_abc1}, and~\eqref{eq:lambda_abc2}--\eqref{eq:lambda_2} resolve to
\begin{subequations}\label{eq:windingsdq}
\begin{align}
    \diff{\Lambda'}{t} &= - I' r + E'-\omega \Lambda',\label{eq:lambda_abc1x}\\
     \diff{\lambda_\mathrm{f}}{t} &= - i_\mathrm{f} r_\mathrm{f} + e_\mathrm{f},\\
    \diff{\lambda_{1}}{t} &= - i_{1} r_{1} + e_{1},\\
    \diff{\lambda_{2}}{t} &= - i_{2} r_{2} + e_{2},\\
    \diff{\theta_\mathrm{r}}{t} &= \frac{2}{\eta}\omega,\\
    J\frac{2}{\eta}\diff{\omega}{t} &= T_{\mathrm{m}} - T_{\mathrm{e}} - D_\mathrm{w}\frac{2}{\eta}\omega,\\
    \Lambda' &= \Big(l_\mathrm{ls}+\frac{3}{2}l_{\mathrm{A}}\Big)I' \nonumber \\ &- \frac{3}{2}\big((l_{\mathrm{B}}(I')^* - l_\mathrm{s2}i_\mathrm{2}) - \jmath (l_\mathrm{sf} i_\mathrm{f} + l_\mathrm{s1}i_\mathrm{1})\big),\label{eq:lambda_abc2x}\\
    \lambda_\mathrm{f} &= -\Im(I') + l_\mathrm{ff} i_\mathrm{f} + l_\mathrm{f1} i_\mathrm{1},\label{eq:lambda_fx}\\
    \lambda_\mathrm{1} &= -\Im(I') + l_\mathrm{f1} i_\mathrm{f} + l_\mathrm{11} i_\mathrm{1},\label{eq:lambda_1x}\\
    \lambda_\mathrm{2} &= \Re(I') + l_\mathrm{22} i_\mathrm{2}.\label{eq:lambda_2x}
\end{align}
\end{subequations} 
Figure~\ref{Fig:SG}(b) provides an abstract representation of the model discussed above in the local $\mathrm{dq}$ and global $\mathrm{DQ}$ reference frames. Notably, since this figure does not explicitly depict the stator and field dynamics, and these were the main ones simplified as a result of the application of the reference-frame transformation, we do not see much difference between Fig.~\ref{Fig:SG}(a) and Fig.~\ref{Fig:SG}(b) in terms of illustrated quantities. This is because, the emphasis of these figures is the voltage regulator, frequency governor, and rotor dynamics. Changes are notable in the network-facing controlled-voltage  interface in the global $\mathrm{DQ}$ reference frame.

\noindent \subsubsection*{Steady-state Model} By setting all the derivative terms in~\eqref{eq:windings},~\eqref{eq:exciter}, and~\eqref{eq:governor} to zero, and noting that $e_{1}=e_{2}=0$ since damper windings have no external excitation, we obtain:
\begin{subequations}
\begin{align}
        T_{\mathrm{e}} =&\ \frac{1}{{\omega}_\mathrm{s}}P^\star - \frac{\kappa_{\mathrm{p}}}{R^\mathrm{d}\omega_{\mathrm{s}}}({\omega}_\mathrm{ss}-{\omega_\mathrm{s}}) - D_\mathrm{w}\frac{2}{\eta}{\omega}_\mathrm{ss}, \label{eq:sg-steady1} \\
        E' =&\ \jmath\omega_\mathrm{ss}\frac{3 l_\mathrm{sf}\kappa_\mathrm{a}}{2r_\mathrm{f}\kappa_\mathrm{e}}\Big(|E^\star|-|E'|-\frac{3\kappa_\mathrm{c}}{2|E'|}Q\Big) + I' r \nonumber\\&+ \omega_\mathrm{ss}\Big(l_\mathrm{ls}+\frac{3}{2}l_{\mathrm{A}}\Big)I' - \frac{3}{2}\omega_\mathrm{ss} l_{\mathrm{B}}(I')^*. \label{eq:sg-steady2}
\end{align}
\end{subequations}
In what follows, we will apply reasonable assumptions that align~\eqref{eq:sg-steady1}--\eqref{eq:sg-steady2} with the popular $PV$-bus representation of synchronous generators in steady-state power-flow models.

Now, suppose that ${D_\mathrm{w}}\approx0$, $\kappa_{\mathrm{p}}\approx 1$, and the steady-state frequency ${\omega}_\mathrm{ss}$ is such that $$P = {\omega}_\mathrm{ss}T_{\mathrm{e}}\approx {\omega}_\mathrm{s}T_{\mathrm{e}},$$ then we have that
\begin{align}
    P \approx&\ P^\star - \frac{1}{R^\mathrm{d}}({\omega}_\mathrm{ss}-{\omega_\mathrm{s}}).\label{eq:Pss}
\end{align}

Now, suppose $r\approx0$, $l_\mathrm{ls}\approx0$, and $l_{\mathrm{A}}-l_{\mathrm{B}}\approx0$, then, $\Re{(E')} = 0$, and we get the following from~\eqref{eq:sg-steady2}:
\begin{align}
    |\Im{(E')}| = |E'|=& \Bigl|\omega_\mathrm{ss}\frac{3 l_\mathrm{sf}\kappa_\mathrm{a}}{2r_\mathrm{f}\kappa_\mathrm{e}}\Big(|E^\star|-|E'|-\frac{3\kappa_\mathrm{c}}{2|E'|}Q\Big) \nonumber\\&+ \frac{3}{2}\omega_\mathrm{ss}(l_{\mathrm{A}}+l_{\mathrm{B}})\Im{(I')}\Bigr|.\label{eq:sg-steady3}
\end{align} Using the fact that $\Im{(I')}=\frac{3Q}{2|E'|}$, it follows that 
\begin{align}
    |E'|^2 =& \Bigl|\frac{3 l_\mathrm{sf}\kappa_\mathrm{a}\omega_\mathrm{ss}}{2r_\mathrm{f}\kappa_\mathrm{e}+3 l_\mathrm{sf}\kappa_\mathrm{a}\omega_\mathrm{ss}}|E^\star|{|E'|} \nonumber\\&- \frac{{9}\big(l_\mathrm{sf}\kappa_\mathrm{a}{\kappa_\mathrm{c}}-r_\mathrm{f}\kappa_\mathrm{e}(l_{\mathrm{A}}+l_{\mathrm{B}})\big)\omega_\mathrm{ss}}{4r_\mathrm{f}\kappa_\mathrm{e}+6 l_\mathrm{sf}\kappa_\mathrm{a}\omega_\mathrm{ss}}{Q}\Bigr|.\label{eq:sg-steady4}
\end{align} 
Now, considering that the value of $r_\mathrm{f}$ is small and in the regime $\kappa_\mathrm{c} \to 0$, we can show~\eqref{eq:sg-steady4} simplifies to $|E'| \approx |E^\star|$. In terms of sinusoidal steady-state phasor representation, we equivalently write
\begin{align}
    |\overline E| \approx |\overline E^\star|, \label{eq:Ess}
\end{align}
where $|\overline E^\star| = \sfrac{\sqrt{2}}{3} |E^\star|$.
Figure~\ref{Fig:SG}(c) illustrates the (approximate) steady-state phasor model of the generator as described mathematically in~\eqref{eq:Ess} and~\eqref{eq:Pss}. This aligns, in principle, with the popular voltage-behind-reactance model(s) for generators (see, e.g.,~\cite{SauerPai-1998-Book}).   
\begin{table*}
\centering
    \caption{Generic GFM primary-control model and parametric assumptions to recover droop, VSM, and dVOC. See~\cite{Ajala22HiCSS} for detailed description of variables.}
    \renewcommand\arraystretch{1.3}
     \begin{tabular}{lcccccc}
        \toprule
        \midrule
        Type & $\tau_{\mathrm f}$ & $\tau_{\mathrm e}$ & $\kappa_{\mathrm{f}}$ &  $\kappa_{\mathrm e}$ & $\kappa_{\mathrm d}$ &  $f_{\mathrm e}(|E'|,|E^\star|)$  \\
        \midrule
        \midrule
        dVOC  & 0   & 1 & $\frac{k_{1}}{|E'|^{2}}$   & $\frac{k_{1}}{|E'|}$ &  0 &  $k_{2}|E'|(|E^\star|^2-|E'|^2)$ \\
        \cmidrule(r){1-1}\cmidrule(l){2-7}
        droop  & 0 & 0 & $M^{\mathrm P}$   & $M^{\mathrm Q}$ &  0 &  $|E^\star|- |E'|$  \\
        \cmidrule(r){1-1}\cmidrule(l){2-7}
        VSM  & $M^{\mathrm P} M$   & 0 & $M^{\mathrm P}$ & $M^{\mathrm Q}$     & $M^{\mathrm P}D_{\mathrm d}$  & $|E^\star|- |E'|$ \\
        \midrule
        \bottomrule
    \end{tabular}
    \label{tab:UGFM}
\end{table*}
\subsection{Grid-following (GFL) Inverter Model} \label{sec:GFL}
The GFL inverter employs a synchronous-reference-frame~(SRF) phase-locked loop~(PLL) for grid synchronization and proportional integral~(PI) current- and power-control loops for reference tracking~\cite{Iravani_Book10}. The PI controllers operate on signals transformed to the local $\mathrm{dq}$ reference frame. The voltage, $E'$, and GFL frequency, $\omega$, are given by: \begin{subequations}
\begin{align}
    E' &= \left(K^{\mathrm p}_{I} \diff{\Gamma'_I}{t} + K^{\mathrm i}_{I} \Gamma'_I\right), \label{eq:GFL-E}\\
    \omega &=\omega_\mrs + K^{\mathrm p}_{PLL}  \diff{\Phi'}{t}+ K^{\mathrm i}_{PLL} \Phi', \label{eq:GFL-Theta}
\end{align}
\end{subequations}
where, $\Gamma'_I$ is the current-controller state, $(K^{\mathrm p}_{I}, K^{\mathrm i}_{I})$ are the corresponding PI gains; $\Phi'$ is the PLL state, and $(K^{\mathrm p}_{PLL}, K^{\mathrm i}_{PLL})$ are the corresponding PI gains. Note that $E'$ is complex valued; magnitude $|E'|$ and phase $\delta'$ pertinent to network-facing representation of the GFL model (as referenced in~\eqref{eq:ea}--\eqref{eq:ec} and~\eqref{eq:E}) can be straightforwardly extracted from~\eqref{eq:GFL-E}. The PI control loops referenced in~\eqref{eq:GFL-E}--\eqref{eq:GFL-Theta} are closed around the measured inverter current, $I'$, and grid voltage, $V'$, such that the grid space phasor, $\overrightarrow{V}$, is aligned with the $\mathrm q$ axis~\cite{DVR2}: 
\begin{subequations}
\begin{align}
    \diff{\Phi'}{t} &= -\Re(V'), \label{eq:GFL-phi} \\
       \diff{\Gamma'_I}{t} &= I'^\star - I', \label{eq:GammaI} 
 \end{align}  
\end{subequations}
where, $V'$ ($=\bGamma \bv \mathrm{e}^{-\jmath \int_{\tau = 0}^{t} \omega(\tau) \mathrm{d}\tau}$) is obtained using~\eqref{eq:phasor-ABC} and~\eqref{eq:dqdef}, and  the current reference, $I'^\star$, is derived from an outer PI control loop closed around the measured inverter power, $S$, and takes the power reference, $S^\star$:
\begin{subequations}
\begin{align} 
    I'^\star &= K^{\mathrm p}_{S} \diff{{\Gamma}'_S}{t} + K^{\mathrm i}_{S}  {\Gamma}'_S, \\
     \diff{{\Gamma}'_S}{t} &= S^\star - S.  \label{eq:GammaS}
\end{align}
\end{subequations}
Above, ${\Gamma}'_S$ is the power-controller state, $(K^{\mathrm p}_{S}, K^{\mathrm i}_{S})$ are the corresponding PI gains. 

Figure~\ref{Fig:GFL}(a) illustrates pertinent aspects of the GFL dynamical model discussed above as an $\mathrm{abc}$ model. We note that this is true-to-form, in the sense that the $\mathrm{dq}$-domain representation of controller dynamics is per practice and not a model-representation artefact. Figure~\ref{Fig:GFL}(b) illustrates the GFL dynamics in a manner that permits integration with a global $\mathrm{DQ}$ model for the network. The major change across the two figures is in the adopted reference-frame transformations.

\noindent \subsubsection*{Steady-state Model} In steady-state, the derivatives of state variables listed above vanish, and all state variables attain a constant value. Algebraic expressions pertinent to steady-state modeling of GFL IBRs can be obtained from \eqref{eq:GammaS} as:
\begin{subequations}
\begin{align}
     P &= P^\star\label{eq:GFL-P}, \\
      Q &= Q^\star\label{eq:GFL-Q}.
\end{align}
\end{subequations}
Therefore, the equivalent current-dependent voltage-source model for GFLs is captured by
\begin{equation}
   \overline{E}=\frac{P^\star+\jmath Q^\star}{3\overline{I}^*}=\frac{S^\star}{3\overline{I}^*}.
\end{equation}
Figure~\ref{Fig:GFL}(c) shows how the above description maps to an equivalent-circuit representation in sinusoidal steady state.

\subsection{Grid-forming Inverter Model} \label{sec:GFM}

The grid-forming inverter behaves much like a synchronous machine in terms of its terminal voltage characteristics and synchronizes with the grid without requiring an explicit PLL. Three popular incarnations of GFM inverter control exist: droop, virtual synchronous machine, and dispatchable virtual oscillator control \cite{Ajala-CDC,Ajala22HiCSS}. The \emph{generic} GFM model discussed below admits all three popular GFM-IBR control dynamics under certain parametric assumptions (see Table~\ref{tab:UGFM}). The GFM controllers (in practical implementations) operate on signals transformed to the local $\mathrm{dq}$ reference frame and generate the terminal voltage, $E'$, at the GFM frequency, $\omega$. The controller dynamics are captured by: 
\begin{subequations}
\begin{align}
   \tau_{\mathrm e} \diff{|E|'}{t} &= f_{\mathrm e}(E',|E^\star|) - \kappa_{\mathrm e} \Im{( S^{\mathrm f} - S^\star)}, \label{eq:GFM-Vi} \\
     \omega &=\omega_\mrs -\tau_{\mathrm f} \diff{\omega}{t} + \kappa_{\mathrm f}\Re{( S^{\mathrm f} - S^\star)} \notag \\ &+ \kappa_{\mathrm d}\left(K^{\mathrm p}_{{PLL}}\diff{\Phi'}{t} + K^{\mathrm i}_{{PLL}} \Phi'\right), \label{eq:GFM-omegai} 
 \end{align}
\end{subequations}
where, ($\tau_{\mathrm e}$, $\tau_{\mathrm f}$) denote the time constants for the voltage and frequency loops; variable $|E^\star|$ denotes the voltage-magnitude setpoint; the function $f_{\mathrm e}(\cdot,\cdot)$ represents a voltage difference metric; ($\kappa_{\mathrm e}$, $\kappa_{\mathrm f}$) capture the voltage- and frequency-controller coefficients of the primary controller. In the case of the VSM, an additional PLL is employed as an ac-side frequency sensor with the variable $\Phi'$ denoting the PLL's internal state, $(K^{\mathrm p}_{PLL}, K^{\mathrm i}_{PLL})$ correspond to the PI gains, and $\kappa_{\mathrm d}$ captures the VSM damping coefficient; the quantity $S^{\mathrm f}$ represents the low-pass-filtered measurements of active and reactive powers. The dynamics of these are given by: \begin{subequations}
\begin{align}
        \diff{\Phi'}{t} &= -\Re{(E')}, \label{eq:GFM-phi} \\
      \tau_{\mathrm p} \diff{{S}^{\mathrm f}}{t} &= S- S^{\mathrm f},  \label{eq:GFM-power} 
\end{align}
\end{subequations}
where $\tau_{\mathrm p}$ denotes the power-filter time constant. As with the GFL model, $E'$ is complex valued; but the GFM dynamics only modulate the magnitude $|E'|$. Implicit in this is the fact that phase $\delta'=0$ for all GFM controllers.

Figure~\ref{Fig:GFM}(a) illustrates the GFM dynamical model discussed above in the $\mathrm{abc}$ frame that is true-to-form. The representation of controller dynamics in the $\mathrm{dq}$ frame is per practice. Similar to the GFL case, Fig.~\ref{Fig:GFM}(b) leverages appropriate reference-frame transformations and casts the GFM model in a manner that permits integration with a global $\mathrm{DQ}$ network model. 

\noindent \subsubsection*{Steady-state Model} In steady state, the derivatives in the dynamic equations vanish and all state variables attain a constant value. The algebraic expressions pertinent to steady-state modeling of the GFM are
\begin{subequations}
\begin{align}
    P &= P^\star{-}\frac{1}{M^\mathrm{P}}(\omega_\mrss-\omega_\mathrm{s}),\\
    Q &\approx Q^\star{-}\frac{1}{M^\mathrm{Q}}(|E'|{-}|E^\star|) = Q^\star{-}\frac{1}{\overline M^\mathrm{Q}}(|\overline E|{-}|\overline E^\star|),
\end{align}
\end{subequations}
where, $|\overline E^\star| = \sfrac{\sqrt{2}}{3} |E^\star|$, and $\overline M^\mathrm{Q} = \sfrac{\sqrt{2}}{3}M^\mathrm{Q}$. The relationship between the active power, $P$, and frequency, $\omega$, follows the familiar droop characteristics for all three types of GFMs. On the other hand, the relationship between reactive power, $Q$, and voltage magnitude, $|E'|$, is (approximately) linear droop with tuning in the case of dVOC. The equivalent current-dependent voltage-source model for GFMs is captured by
\begin{equation}
   \overline{E}=\frac{P+\jmath Q}{3\overline{I}^*}.
\end{equation}
Figure~\ref{Fig:GFM}(c) shows how the above description maps to an equivalent-circuit representation for sinusoidal steady state.
\section{Offshoots, Inferences, and Corollaries}\label{sec:ICO}
In this section, we overview approximate models~$(\mathcal{M}2')$, $(\mathcal{M}3')$; comment on how power-flow equations that are ubiquitous with the steady-state modeling of power grids are aligned with the descriptions $(\mathcal{M}3)$, $(\mathcal{M}3')$; and finally, offer a closed-form approximate expression for steady-state frequency, $\omega_\mathrm{ss}$, which is central to realizing the steady-state models $(\mathcal{M}3)$, $(\mathcal{M}3')$. 

\subsection{Approximate Models with Dubious Justification} \label{sec:Approx}
The previous sections provide principled  dynamic and steady-state network models with accompanying abstract-resource representations. In this section, we overview two approximate network-resource system models that are prevalent in the literature, but warrant careful scrutiny. Notably, we will find these are challenging to formalize, even with the rigorous modeling foundations we have provided in this work. 
\subsubsection*{DAE Model~$(\mathcal{M}2')$} In traditional power system dynamic analysis, the line dynamics of power networks are often replaced by steady-state algebraic relations, while retaining the resource-interface dynamics~\cite{SauerPai-1998-Book}. Moreover, the steady-state network modeling is carried out assuming nominal frequency $\omega_\mrs$. Stating in terms of the setup under consideration, such practice advocates the following DAE model:
\begin{align} \label{eq:DAE} \tag{$\mathcal{M}2'$}
\begin{split} 
    \diff{\bchi_n}{t}&=\bof_n(\bchi_n,\by_n),~\forall~n\in\mcN\\
    \bzero&=\bg_n(\bchi_n,\by_n),~\forall~n\in\mcN\\
    \bL^\mcI \diff{\bI}{t} &= \bE {-} \bV -\bR^\mcI\bI- \jmath \omega_\mathrm{s} \bL^\mcI \bI,   \\
    \bI&=\bY^{\subset}(\omega_\mathrm{s})\bV.
\end{split}
\end{align}
In conventional synchronous generator-dominated power systems, such models typically stem from identifying time-scale separation between resource and network dynamics. However, in power grids with diverse resources, one needs to re-examine the validity of such simplifications. Two potential adverse effects of inapt DAE-simplifications include: i) inability to capture dynamic interactions between network $RL$ lines and resource dynamics; and ii) inaccuracy from frequency-agnostic network modeling across timescales.

\subsubsection*{Synchronous Steady-state Phasor Model~$(\mathcal{M}3')$}
Assessing the sinusoidal steady-state operating point for a network hosting diverse resources is a problem of great interest. For the setup considered in this work, the sinusoidal steady state is characterized by the collection of resource models~\eqref{eq:IBRss} and network model~\eqref{eq:RLss}. However, some studies retain the accurate resource steady-state models \eqref{eq:IBRss} but omit the steady-state frequency dependence from \eqref{eq:RLss} (or equivalently \eqref{eq:IYV}). The resulting setup reads as:
\begin{align} \label{eq:SynSS} \tag{$\mathcal{M}3'$}
\begin{split} 
    &\overline{\bof}_n(\overline{E}_n,|\overline{E}_n^\star|, \overline{I}_n,P^\star_n,Q^\star_n,\omega_\mathrm{ss})=\bzero,\\
    &\overline{\bI}=\bY(\omega_\mathrm{s})\overline{\bE}.
\end{split}
\end{align}

\subsection{Power Flow}\label{sec:PF}
In its most commonly referenced form, the classical power flow problem is a steady-state analysis problem for synchronous generators. It aspires to solve for network voltages and phase angles as a function of active- and reactive-power injections. Curiously, the problem is always presented in a manner that precludes its connections to underlying dynamics of resources, obscures the role of tertiary and secondary control on the formulation, and with poorly justified constitutive elements such as the slack/swing bus and angle references. (That said, some recent works have attempted to clarify longstanding confusion on some of these aspects~\cite{Dhople-slack-2020,Baros-2021}.) Furthermore, problem formulations have not caught up to systematically acknowledge inverter dynamics, with most efforts attempting to fit in inverters as part of already existing formulations. 

This section puts forth a novel rendition of the steady-state power-flow problem that acknowledges the steady-state frequency to be unknown and suitably incorporates resource terminal behavior for generators and inverters. We will start with the steady-state network model~\eqref{eq:RLss}, equivalently represented by \eqref{eq:IYV}, and instantiate the resource-specific models for \eqref{eq:IBRss}. Subsequently, we will elucidate the problem (un)knowns and finally contrast the formulated power flow instance against classical setups. Perhaps the most important contribution of this formulation will be to clearly connect phase angles, voltage magnitudes (and frequency) as they are solved for in steady state with the originating dynamics. 

Considering power flow instances typically present the governing equations in terms of the nodal (re)active power injections, let us compute, using \eqref{eq:IYV}, the power injections as
\begin{equation}
    \bP+\jmath \bQ =3\overline{\bE}\circ(\bY(\omega_\mrss)^*\overline{\bE}^*).
\end{equation}
Let us separate the real and imaginary parts of the admittance matrix as  $\bY(\omega_\mrss)=\bG(\omega_\mrss)+\jmath \bB(\omega_\mrss)$. With the introduced notation, the power injections at node $n\in\mcN$ are given by
\begin{subequations}\label{eq:PF52}
\begin{align}
    P_n&=3|\overline{E}_n|\sum_{k=1}^N|\overline{E}_k|(G_{nk}(\omega_\mathrm{ss})\cos\overline{\delta}_{nk} \nonumber \\ &\quad \quad \quad \quad \quad \quad +B_{nk}(\omega_\mathrm{ss})\sin\overline{\delta}_{nk}),\label{eq:nodePFP}\\
    Q_n&=3|\overline{E}_n|\sum_{k=1}^N|\overline{E}_k|(G_{nk}(\omega_\mathrm{ss})\sin\overline{\delta}_{nk} \nonumber \\
    &\quad \quad \quad \quad \quad \quad-B_{nk}(\omega_\mathrm{ss})\cos\overline{\delta}_{nk}),\label{eq:nodePFQ}
\end{align}
\end{subequations}
where, the phase difference $\overline{\delta}_n-\overline{\delta}_k$ is written as $\overline{\delta}_{nk}$, and the $(n,k)$-th entry of frequency-dependent matrices $(\bG(\omega_\mrss), \bB(\omega_\mrss))$ are denoted as $(G_{nk}(\omega_\mrss),B_{nk}(\omega_\mrss))$.

Next, we bring forth the steady-state behavior of the different resources outlined in Section~\ref{sec:Resource} to specify the power injections in~\eqref{eq:nodePFP}--\eqref{eq:nodePFQ}. In particular, based on the type of resource connected at node $n\in\mcN$, the governing equations to be considered are
\begin{subequations}\label{eq:resourcePF}
\begin{align}
    \textrm{SG:}\quad P_n&=P_n^\star-\frac{1}{R^\mathrm{d}_n}(\omega_\mrss-\omega_s)\\
    \overline{E}_n& =  |\overline{E}_n^\star| \\
    \textrm{GFL:}\quad P_n&=P_n^\star\\
    Q_n&=Q_n^\star\\
    \textrm{GFM:}\quad P_n&=P_n^\star-\frac{1}{M^\mathrm{P}_n}(\omega_\mrss-\omega_s)\\
    Q_n&=Q_n^\star-\frac{1}{\overline{M}^{\mathrm Q}_n}(|\overline{E}_n|-|\overline{E}_n^\star|).
\end{align}
\end{subequations} 

While solving for the steady-state operating point of the integrated system, besides the network parameters, one would be given: (re)active power setpoints $(P_n^\star,Q_n^\star)$ for GFLs and GFMs, active power setpoints $P_n^\star$ for SGs, and voltage magnitude setpoints $|E_n^\star|$ for GFMs and SGs. One would thus seek to solve for the $4N+1$ real-valued variables in $(\bP,\bQ,|\overline{\bE}|,\overline{\bdelta},\omega_\mrss)$. Based on resource-type,~\eqref{eq:resourcePF} offers $2$ equations per node. Combining these with the $2N$ equations in \eqref{eq:nodePFP}-\eqref{eq:nodePFQ}, one obtains $4N$ equations, with the final equation needed for well-posedness being given by setting a reference bus angle, that is
\begin{equation}
    \overline{\delta}_r=0,
\end{equation}
for some bus $r\in\mcN$. The presented power-flow formulation does not explicitly feature the network-node voltages $\overline{\bV}$ and line currents $\overline{\bF}$. However, these can be recovered a posteriori via the first two equations of \eqref{eq:RLss}.

We next compare with traditional power-flow setups in power transmission systems. Albeit equations~\eqref{eq:nodePFP}-\eqref{eq:nodePFQ} show up in tradition power-flow settings, the frequency dependence in $(\bG(\omega_\mrss),\bB(\omega_\mrss))$ is omitted by approximately modeling the network at nominal frequency $\omega_\mrs$. Next, traditional power-flow instances primarily feature two types of buses: constant (re)active power injection--PQ buses; and constant active power and voltage magnitude--PV buses. In the resource models of \eqref{eq:resourcePF}, if one omits all the droop offsets, the GFL and GFM models would qualify as PQ buses, while SGs would serve as PV buses. 

\subsection{Steady-state Frequency} \label{sec:omegass}
The power-flow model put forth in Section~\ref{sec:PF}, once solved for given power and voltage setpoints, yields the steady-state operating point that features the steady-state frequency $\omega_\mrss$. Albeit precise, the formulated problem involves tedious solving of non-linear system of algebraic equations where the resulting complexity grows with the network dimension. Oftentimes, network operators are particularly interested in the value of $\omega_\mrss$ and seek an efficient way for computing it. In this regard, one can observe from~\eqref{eq:PF52}--\eqref{eq:resourcePF} that equations governing $\omega_\mrss$ are not separable. Thus any approach that promises to bypass the complexity of solving \eqref{eq:PF52}--\eqref{eq:resourcePF} can, at best, find an approximation to $\omega_\mrss$. We briefly present one such approach next.

Let us invoke the assumption of lossless network, frequently adopted for high-voltage power systems. The assumption sets $\bG=\bzero$, which on using the property of $\sin()$ being an even function in \eqref{eq:nodePFP} yields $\sum_{n=1}^N P_n=0$. We will use the aforementioned in computing $\omega_\mrss$ from the equations governing active power in \eqref{eq:resourcePF}. To do so, let us define subsets of $\mcN$ based on the resource type as $(\mcN_L,\mcN_M,\mcN_S)$ hosting the GFL IBRs, GFM IBRs, and SGs, respectively; where $\mcN=\mcN_L\cup\mcN_M\cup\mcN_S$. The zero-sum property of steady-state active power injections then yields 
\begin{align*}
    0&=\sum_{n=1}^N P_n=\sum_{n\in\mcN_L}P_n+\sum_{n\in\mcN_M}P_n+\sum_{n\in\mcN_S}P_n\\
    &=\sum_{n\in\mcN_S}\left({P}_n^\star-\frac{1}{R^\mathrm{d}_n}(\omega_\mrss-\omega_\mrs)\right) + \sum_{n\in\mcN_L}{P}_n^\star  \\&+\sum_{n\in\mcN_M}\left({P}_n^\star-\frac{1}{M_n^\mathrm{P}}(\omega_\mrss-\omega_\mrs)\right),
\end{align*}
where the second equality follows from~\eqref{eq:resourcePF}. Solving the above provides
\begin{equation}
    \omega_\mrss=\omega_\mrs+\frac{\sum_{n\in\mcN}P_n^\star}{\sum_{n\in\mcN_M}\sfrac{1}{M_n^\mathrm{P}}+\sum_{n\in\mcN_S}\sfrac{1}{R^\mathrm{d}_n}}.
\end{equation}


\section{Concluding Remarks}\label{sec:conclude}
This paper outlined integrated system models for generators and inverters in different reference frames and timescales. A dependent voltage-source resource model was developed for the generators and inverters, and suitable instantiations were highlighted for $\mathrm{abc}$, direct-quadrature, and steady-state models. Connections to ubiquitous models including DAE and power-flow were outlined. A clear link between network facing quantities was retained across model types.  






\balance
\bibliographystyle{IEEEtran}
%

\bibliography{myabrv,IREP}




\end{document}